\documentclass[superscriptaddress,aps,prl,reprint,twocolumn,amsmath,amssymb]{revtex4-2}
\usepackage{graphicx}
\usepackage{hyperref}
\usepackage{amssymb}
\usepackage{slashed}
\usepackage{dcolumn}
\usepackage{amsmath}
\usepackage{bm}
\usepackage{colordvi}
\usepackage{algorithm}
\usepackage{algpseudocode}
\usepackage{multirow}
\usepackage{titlesec}
\usepackage{newtxtext,newtxmath}
\usepackage{dsfont}
\usepackage{xcolor}
\usepackage{mathbbol}

\usepackage{hyperref}
\hypersetup{
    colorlinks=true,
    linkcolor=blue,
    citecolor=blue,     
    urlcolor=blue,
}

\allowdisplaybreaks

\usepackage{mathrsfs}
\makeatletter

\newcommand{\Rmnum}[1]{\expandafter\@slowromancap\romannumeral #1@}
\makeatother

\begin{document}
	\preprint{APS/123-QED}
	\title{Excitonic-Superconducting Coexistence and Emergent Nematic Superconductivity Driven by Spontaneous Symmetry Breaking}

\author{Fei Yang}
\thanks{These authors contributed equally to this work.}

\affiliation{Department of Physics, The Hong Kong University of Science and Technology, Clear Water Bay, Kowloon, Hong Kong SAR, China}

\author{Ruigang Li}
\thanks{These authors contributed equally to this work.}

\affiliation{Department of Physics, National University of Singapore, Singapore 117551, Republic of Singapore}

\author{Junwei Liu}
\email{liuj@ust.hk}
\affiliation{Department of Physics, The Hong Kong University of Science and Technology, Clear Water Bay, Kowloon, Hong Kong SAR, China}

\author{Binghai Yan}
\email{binghai.yan@psu.edu}
\affiliation{Department of Physics, The Pennsylvania State University, University Park, PA 16802, USA}
\affiliation{Center for Theory of Emergent Quantum Matter, The Pennsylvania State University, University Park, PA 16802, USA}

\begin{abstract}
Excitonic insulating (EI) and superconducting (SC) orders are generally regarded as mutually exclusive electronic instabilities. Within a self-consistent microscopic theory, we study electronic systems hosting an EI phase in the presence of SC pairing and show that an intrinsic mismatch between electron and hole Fermi surfaces fundamentally reshapes this competition. This mismatch stabilizes FFLO-like electron-hole pairing and  drives spontaneous symmetry breaking of the EI state. The resulting symmetry breaking reconstructs the pairing phase space for SC and EI state, such that different regions of the Fermi surface complementarily support either EI or SC correlations, leading to a natural coexistence of the two orders. Notably, the emergent SC state consequently breaks rotational symmetry and develops intrinsic nematic superconductivity, even in the absence of explicit symmetry-breaking fields (such as magnetic fields, spin-orbit coupling, or bare band-structure anisotropy). Our results suggest that candidate materials such as monolayer 1T$'$-MoTe$_2$ and the square-net semimetal NaAlSi may provide promising platforms  for observing this phenomenon. More broadly, these findings reveal a unique  mechanism by which competing many-body orders generate electronic nematicity, suggesting a broader route toward spontaneous anisotropic electronic states in correlated quantum materials.
\end{abstract}

\maketitle

{\sl Introduction.---}Over the past few decades, advances in materials synthesis and characterization have enabled the discovery of a broad class of correlated quantum materials, where multiple electronic orders and their interplay have attracted significant attention, as the competition and coexistence of these collective orders often give rise to emergent phenomena and unconventional many-body ground states. Among these, superconducting (SC)~\cite{PhysRev.108.1175,schrieffer1964theory,tinkham2004introduction,abrikosov2012methods,mahan2013many} and excitonic insulating (EI)~\cite{PhysRev.158.462,PhysRevLett.19.439,RevModPhys.40.755} orders represent two distinct pairing instabilities that admit a unified description in terms of fermionic pairing.

The SC state arises from electron-electron pairing near the Fermi surface, mediated by an effective attractive interaction~\cite{PhysRev.108.1175,schrieffer1964theory,tinkham2004introduction,abrikosov2012methods,mahan2013many}, typically originating from bosonic fluctuations such as phonons, leading to the formation of Cooper pairs. In contrast, the EI state is driven by Coulomb interactions in systems with overlapping or nearly touching conduction and valence bands~\cite{PhysRev.158.462,PhysRevLett.19.439,RevModPhys.40.755}, where electrons and holes bind into excitons that subsequently condense and open an electronic gap. Such small- or negative-gap conditions, ensuring the simultaneous presence of electrons and holes, have been realized in a variety of systems, including narrow-gap semiconductors~\cite{PhysRevB.104.085133,Grudinina2023,Du2017,Lu2017}, transition metal dichalcogenides~\cite{PhysRevLett.134.066602,Bi2021,Jia2022,Sun2022,PhysRevLett.125.046801,Varsano2020,x6k9-wgk9}, moir\'e heterostructures~\cite{Guo2023,Tran2019,Regan2020,PhysRevLett.118.147401,PhysRevLett.121.026402}, and square-net semimetals~\cite{qfqm-qp82}. Interestingly, several of these systems, including monolayer 1T$'$-WTe$_2$~\cite{Song2024,PhysRevResearch.7.013224,b6vp-zt8z,Fatemi2018,Sajadi2018}, 1T$'$-MoTe$_2$~\cite{x6k9-wgk9,Rhodes2021,px77-3gg1}, and NaAlSi~\cite{qfqm-qp82,PhysRevB.81.245114,Yamada2021,Zhong2024}, also exhibit superconductivity, often with strong electrostatic-gate tunability~\cite{Song2024,PhysRevResearch.7.013224,b6vp-zt8z,Fatemi2018,Sajadi2018,Rhodes2021}. This  enables a continuous control of carrier density, providing  promising platforms for investigating the competition and interplay between SC and EI orders.

Conventional theoretical study~\cite{PhysRevB.71.224502} predicts that SC and EI orders are mutually exclusive in clean systems when both pairing instabilities are present, as they compete for the same low-energy electronic degrees of freedom on the Fermi surface. Then, the development of one order suppresses the other, with the system favoring a single dominant instability that minimizes the thermodynamic potential~\cite{PhysRevB.71.224502}. This conclusion, however, relies on the assumption of near-perfect electron-hole band matching inherent to early EI  theories~\cite{PhysRev.158.462,PhysRevLett.19.439,RevModPhys.40.755,PhysRevB.71.224502}. In many relevant material systems, electron and hole Fermi surfaces are not perfectly matched due to the band-structure asymmetry, carrier doping, or interaction-driven renormalization. Such Fermi surface mismatch can qualitatively modify the nature of electron-hole pairing and may favor excitonic condensates with finite center-of-mass (CM) momentum~\cite{PhysRevLett.134.066602}, analogous to Fulde-Ferrell-Larkin-Ovchinnikov (FFLO) pairing states in superconductors~\cite{larkin1965zh,fulde1964superconductivity}. Then, the EI state can be viewed as a generalized density-wave order in the particle-hole channel, analogous to conventional charge-density waves~\cite{PhysRevB.97.224423,PhysRevLett.35.120,gruner1988dynamics,gruner1985charge}, but driven primarily by Coulomb attraction rather than Fermi surface nesting. Consistent with this picture, increasing experimental evidence suggests that EI order is often intertwined with density-wave instabilities. In several candidate materials (e.g, monolayer 1T$'$-MoTe$_2$~\cite{x6k9-wgk9,px77-3gg1} and NaAlSi~\cite{qfqm-qp82}), spatially modulated electronic states have been  observed, indicating finite-momentum exciton condensation and the associated spontaneous breaking of translational symmetry.

Concerning this possibility, here we show within a self-consistent microscopic  framework that electron-hole Fermi-surface mismatch fundamentally reshapes the competition between EI and SC orders. It destabilizes the conventional zero-momentum excitonic condensate~\cite{PhysRev.158.462,PhysRevLett.19.439,RevModPhys.40.755,PhysRevB.71.224502} and favors spontaneous symmetry breaking of the EI state, stabilizing FFLO-like finite-momentum electron-hole pairing. This finite-momentum EI order reconstructs the pairing phase space, allowing EI pairing to develop on selected portions of the Fermi surface while leaving the remaining regions available for SC pairing, enabling the coexistence of SC and EI orders. The resulting SC state  breaks rotational symmetry, giving rise to an emergent spontaneous nematic superconductivity, even in the absence of explicit symmetry-breaking fields.

\begin{figure}[htb]
  {\includegraphics[width=7.5cm]{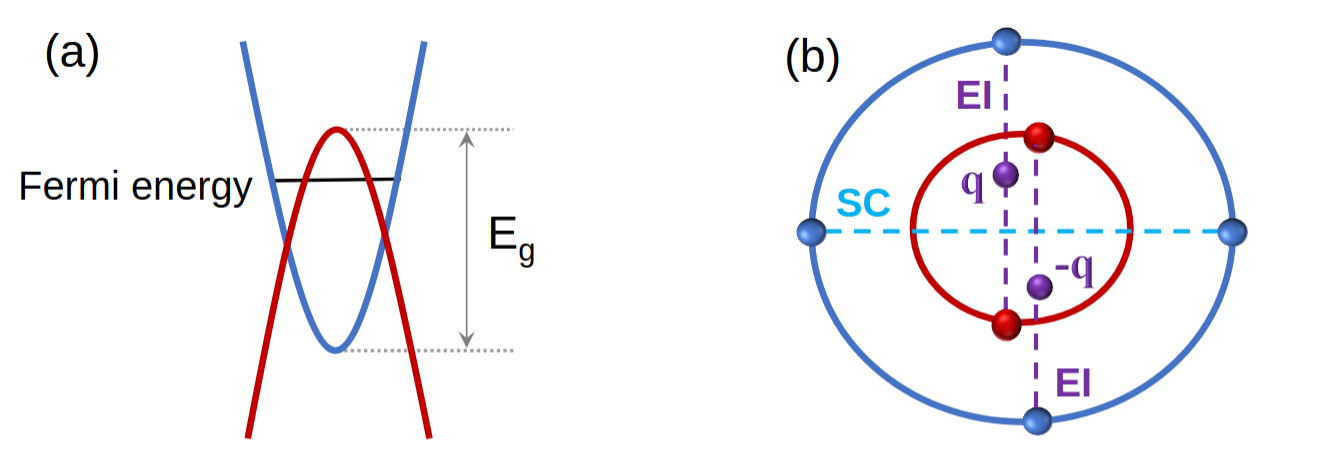}}
\caption{(a) Bare band structure showing the conduction and valence bands with a negative band gap $E_g$. 
(b) Schematic illustration of SC and EI pairing on mismatched Fermi surfaces. 
The purple dashed lines denote electron-hole EI pairings. 
Due to the electron-hole Fermi-surface mismatch, the EI pairing spontaneously acquires finite CM momenta $\pm{\bf q}$, forming an LO-like excitonic state composed of two time-reversal-related pairing components. 
Here, ${\bf q}$ is chosen along the $y$ direction, reflecting the spontaneous rotational-symmetry breaking of the underlying isotropic model. 
As a result, conduction-band states with momenta perpendicular to $\pm{\bf q}$ do not participate in EI pairing, leaving parts of the Fermi surface available for conventional zero-momentum SC pairing (blue dashed curve).
}
\label{FIG1}
\end{figure}

{\sl Model.---}We consider a two-band model consisting of an electron
(conduction) band centered at ${\bf K}_c$ and a hole
(valence) band centered at ${\bf K}_v$. When momenta are measured relative to the respective band extrema, ${\bf K}_c$ and ${\bf K}_v$
do not explicitly enter the band dispersions
or group velocities, and hence do not directly
affect the quasiparticle spectrum, the resulting
gap equations, or the transport properties
within the conventional theoretical framework. Consequently, within this low-energy description,
the valley separation $\Delta{\bf K}={\bf K}_c-{\bf K}_v$
can be absorbed into a momentum-folding representation
that maps the two band extrema onto a common
momentum center. The Hamiltonian, as illustrated in Fig.~\ref{FIG1}(a),
is given by
\begin{eqnarray}
H\!&=&\!\sum_{{\bf k},s}(
\xi_{c,\mathbf k} c^\dagger_{{\bf k}s} c_{{\bf k}s}
\!+\!\xi_{v,\mathbf k} v^\dagger_{{\bf k}s} v_{{\bf k}s})
\!-\!g_{\rm sc}\sum_{\bf k,k'}
c^\dagger_{{\bf k}\uparrow}
c^\dagger_{{-\bf k}\downarrow}
c_{{-\bf k'}\downarrow}
c_{{\bf k'}\uparrow}
\nonumber\\
&&\mbox{}-\!\!\sum_{{\bf kk'q'}ss'}
V_{ss'}({\bf k}\!-\!{\bf k'})
c^\dagger_{{\bf k},s} v_{{\bf k+q'},s'}
v^\dagger_{{\bf k'+q'},s'} c_{{\bf k'},s}.
\end{eqnarray}
Here $c^{\dagger}_{{\bf k}s}$ and
$v^{\dagger}_{{\bf k}s}$
($c_{{\bf k}s}$ and $v_{{\bf k}s}$)
denote fermionic creation (annihilation) operators
in the conduction and valence bands, respectively,
with $\xi_{c,{\bf k}}$ and $\xi_{v,{\bf k}}$
the corresponding band dispersions.
The momentum ${\bf k}$ is measured relative to
the respective band extrema ${\bf K}_c$ and
${\bf K}_v$ in the folded representation.
The interband Coulomb interaction
$V_{ss'}({\bf k}-{\bf k}')$
drives excitonic electron-hole
pairing~\cite{PhysRev.158.462,PhysRevLett.19.439,
RevModPhys.40.755,Bi2021,PhysRevB.71.224502}. {Consistent with experimental observations in related
materials~\cite{Song2024,PhysRevResearch.7.013224,
b6vp-zt8z,Fatemi2018,Sajadi2018,x6k9-wgk9,
Rhodes2021,px77-3gg1,qfqm-qp82,
PhysRevB.81.245114,Yamada2021,Zhong2024}, superconductivity is assumed to originate exclusively
from the conduction band via an effective
attractive interaction $g_{\rm sc}$ in the $s$-wave
channel, whereas the valence band remains non-SC. The adopted pocket-centered dispersion satisfies $\xi_{c,{\bf k},\uparrow}=\xi_{c,-{\bf k},\downarrow}$, allowing pairing without a relative momentum mismatch in this channel. Accordingly, we  consider  opposite-spin pairing between electrons within the same electron pocket, $({\bf K}_c+{\bf k},\uparrow)$ and $({\bf K}_c-{\bf k},\downarrow)$~\cite{ppi}, consistent with recent experimental observations and studies supporting intravalley SC pairing (largely favored by the underlying pairing interaction) in related materials~\cite{x6k9-wgk9,px77-3gg1}. The pairing thus has zero momentum relative to the pocket center, while its total crystal momentum is $2{\bf K}_c$.}

Nevertheless, in the presence of Fermi-surface mismatch between electrons and holes, e.g., arising from different effective masses or carrier doping [see Fig.~\ref{FIG1}(b)], the conventional zero-momentum (${\bf q}=0$) excitonic state becomes unstable. Instead, the system favors an FFLO-like finite-${\bf q}$ electron-hole paired state~\cite{yang2018fulde,larkin1965zh,fulde1964superconductivity,Dong_2013,PhysRevA.89.013607,PhysRevLett.89.227002,PhysRevB.76.014522,PhysRevB.75.064511,PhysRevLett.114.110401,rqp1-jtcb}. We formulate the SC and EI instabilities on an equal footing by applying a mean-field treatment to both the particle-particle and particle-hole channels. Following conventional EI theories~\cite{PhysRev.158.462,PhysRevLett.19.439,RevModPhys.40.755,Bi2021,PhysRevB.71.224502}, we restrict our analysis to the spin-conserving channel, as appropriate for the spin-independent Coulomb interaction, and focus on the leading $s$-wave excitonic instability, i.e., $V_{ss'}({\bf k}-{\bf k}') \simeq g_{\rm EI}\delta_{ss'}$. Accordingly, we introduce the SC~\cite{PhysRev.108.1175,schrieffer1964theory,tinkham2004introduction,abrikosov2012methods,mahan2013many} and EI~\cite{PhysRev.158.462,PhysRevLett.19.439,RevModPhys.40.755,Bi2021,PhysRevB.71.224502} order parameters
\begin{equation}
\Delta_{s}
=g_{\rm sc}\sum_{\bf k}
\langle
c_{-{\bf k}\downarrow} c_{{\bf k}\uparrow}
\rangle,\quad
\Delta_{e}
=g_{\rm EI}\sum_{{\bf k},s}
\langle
v^\dagger_{{\bf k}+{\bf q},s}
c_{{\bf k},s}
\rangle.
\end{equation}
The resulting mean-field Hamiltonian in Nambu space~\cite{abrikosov2012methods}, 
\begin{eqnarray}
{\bar H}\!=\!\!\sum_{\bf k}\psi_{\bf k}^\dagger {\scriptscriptstyle 
\,\begin{pmatrix}
\xi_{c,\mathbf k} & -\Delta_{s} & -\Delta_{e} & 0\\
-\Delta_{s}^\ast & -\xi_{c,\mathbf k} & 0 & \Delta_{e}^\ast\\
-\Delta_{e}^\ast & 0 & \xi_{v,{\bf k+q}} & 0\\
0 & \Delta_{e} & 0 & -\xi_{v,{\bf k+q}}
\end{pmatrix}\,}
\psi_{\bf k}\!+\!\frac{|\Delta_{s}|^2}{g_{\rm sc}}\!+\!\frac{|\Delta_{e}|^2}{g_{\rm EI}}, 
\end{eqnarray}
with the field operator $\Psi_{\bf k}^{\dagger}
=(
c_{{\bf k}\uparrow}^{\dagger},
c_{-{\bf k}\downarrow},
v_{{\bf k}+{\bf q},\uparrow}^{\dagger},
v_{-{\bf k}-{\bf q},\downarrow}
)$. 
Diagonalizing the Hamiltonian yields four quasiparticle branches with energies $\pm E_{+}({\bf k})$ and $\pm E_{-}({\bf k})$, written as
\begin{equation}
E_{\pm}^2({\bf k})=\frac{\xi_{c,\mathbf k}^2
+
\xi_{v,{\bf k+q}}^2
+
|\Delta_{s}|^2
+
2|\Delta_{e}|^2\pm \sqrt{D_{\bf k}}}{2},
\end{equation}
with $D_{\bf k}=\big(\xi_{c,\mathbf k}^2-\xi_{v,{\bf k+q}}^2+|\Delta_{s}|^2\big)^2
+
4|\Delta_{e}|^2[
(\xi_{c,\mathbf k}+\xi_{v,{\bf k+q}})^2
+
|\Delta_{s}|^2]$.
The corresponding thermodynamic potential reads
\begin{equation}
\Omega_{\bf q}=\frac{|\Delta_{s}|^2}{g_{\rm sc}}
+
\frac{|\Delta_{e}|^2}{g_{\rm EI}}-
k_BT\sum_{\bf k}
\sum_{\alpha=\pm}
\ln\Big[2\cosh\frac{E_{\alpha}({\bf k})}{2k_BT}\Big].
\end{equation}
The equilibrium state
of the system are obtained by minimizing the free energy with respect to three variational parameters self-consistently: the SC gap $|\Delta_s|$, the excitonic gap $|\Delta_e|$, and the finite ordering wave vector ${\bf q}$ associated with the excitonic density wave, i.e., $
{\partial \Omega_{\bf q}}/{\partial |\Delta_s|}=0$, 
${\partial \Omega_{\bf q}}/{\partial |\Delta_e|}=0$, and $
{\bf q}
=
\arg\min_{\bf q}
\,
\Omega_{\bf q}(|\Delta_s|,|\Delta_e|)$ (See Sec.~SI of the Supplementary Materials for \emph{analytic} expressions of SC and EI gap equations).

The present work considers the more stable, symmetric LO-like excitonic state~\cite{larkin1965zh} [Fig.~\ref{FIG1}(b)]. Accordingly, the momentum states ${\bf k}$ appearing above are restricted to the half of momentum space favorably paired by ${\bf q}$, while the opposite half is paired by the time-reversal-related vector $-{\bf q}$ during momentum summation, thereby preserving the overall time-reversal symmetry of the system (See Sec.~SI~B).

{\sl SC-EI Coexistence.---}To elucidate the central physics, we consider an isotropic parabolic band structure,
\begin{equation}
\xi_{c,\mathbf k} = -\frac{E_g}{2} + \frac{k^2}{2m_c} - \mu,
\qquad
\xi_{v,\mathbf k} = \frac{E_g}{2} - \frac{k^2}{2m_v} - \mu,
\end{equation}
with $E_g$ being the band offset and $\mu$ the chemical potential. We take the effective masses as $m_c=0.42$ and  $m_v = 0.44m_0$. As the underlying model is rotationally symmetric, the direction of $\mathbf q$ is spontaneously selected, and we choose $\mathbf q \parallel \hat y$. In realistic materials, additional sources of anisotropy, such as anisotropic effective masses or a nonzero valley-separation vector between the electron and hole bands, may exist. Such effects can lift the degeneracy between equivalent symmetry directions and stabilize a fixed orientation (see Sec.~SV for  discussion of monolayer 1T$'$-MoTe$_2$ and the square-net semimetal NaAlSi).

\begin{figure}[h]
  {\includegraphics[width=8.6cm]{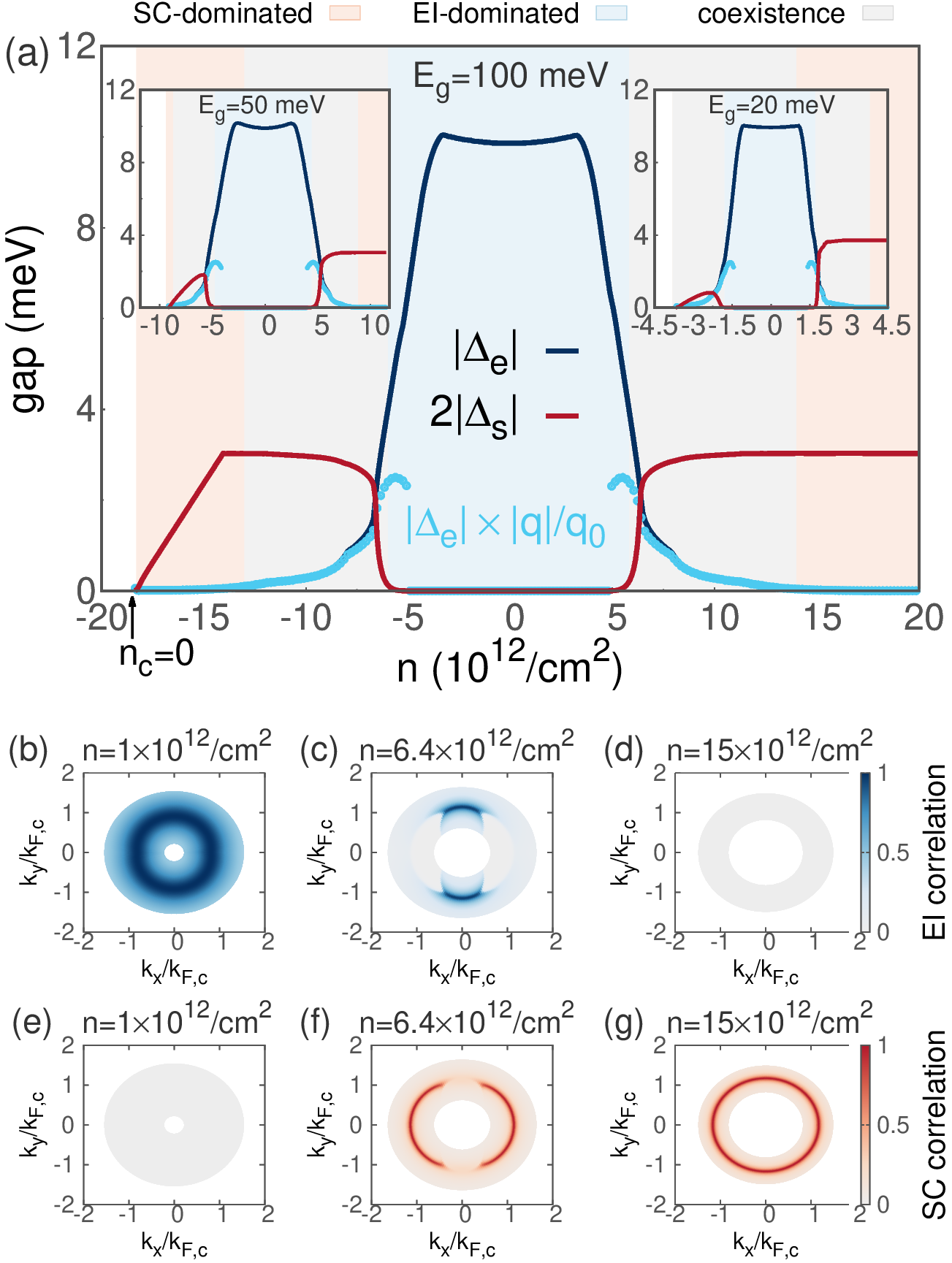}}
\caption{
(a) Carrier-density dependence of the EI gap, SC gap, and the magnitude of {the ordering wave vector} ${\bf q}$ for $E_g=100~\mathrm{meV}$. Here, $n=n_c-n_v$ denotes the net carrier density, and {$q_0=|k_{F,v}-k_{F,c}|$ represents the remaining electron-hole Fermi-surface mismatch.}  Inset: corresponding results for $E_g=50~\mathrm{meV}$ and $20~\mathrm{meV}$. {(b)-(d) and (e)-(g) Momentum-resolved EI and SC correlations [$\rho_{\rm EI}({\bf k})=\langle v_{{\bf k}+{\bf q},s}^{\dagger}c_{{\bf k},s}\rangle$ and 
$\rho_{\rm SC}({\bf k})=\langle c_{{\bf k}\uparrow}c_{-{\bf k}\downarrow}\rangle$] as functions of momentum ${\bf k}$ around the electron Fermi surface, respectively, from full self-consistent calculation. The results are shown at three representative electron-doped carrier densities corresponding to the EI-dominated, EI-SC coexistence, and SC-dominated regimes for $E_g=100~\mathrm{meV}$.}
}
\label{FIG2}
\end{figure}

Motivated by experimental findings~\cite{x6k9-wgk9,px77-3gg1,Song2024,PhysRevResearch.7.013224,b6vp-zt8z,Fatemi2018,Sajadi2018,Rhodes2021}, we consider a regime of dominant excitonic interactions and comparatively weak SC pairing, characterized by $g_{\rm EI} D_c = 0.114$ and $g_{\rm sc} D_c = 0.062$, where $D_c$ is the density of states of the conduction band. All simulations are performed at $T = 0.5~\mathrm{K}$ (see Sec.~SVI for simulation details). The carrier-density dependence ($n=n_c-n_v$) of the EI and SC gaps, as well as the ordering wave vector $q$, is shown in Fig.~\ref{FIG2}(a), together with the corresponding EI and SC correlation strengths (anomalous correlations, See Sec.~SII for details). At charge neutrality ($n\sim0$, i.e., $n_c{\sim}n_v$), {without Fermi-surface mismatch, electrons and holes efficiently form excitonic pairs over the entire Fermi surface. The EI correlations [Fig.~\ref{FIG2}(b)] are therefore isotropically distributed and open a uniform excitonic gap $|\Delta_e|$, which removes the low-energy phase space available for SC pairing. This  competition fully suppresses the SC order ($|\Delta_s|=0$) and stabilizes a $q=0$ excitonic state, as shown in Fig.~\ref{FIG2}(a).} 

Upon carrier doping, as shown in Fig. \ref{FIG2}(a), on either the electron- ($n>0$) or hole-doped ($n<0$) side, the EI gap initially remains insensitive to carrier density, with $q\sim0$, indicating the robustness of the conventional zero-$q$ EI condensate against weak Fermi-surface mismatch. {Beyond a critical doping level, the increasing mismatch between the electron and hole Fermi surfaces drives an abrupt transition toward a finite-${\bf q}$ EI, accompanied by a rapid continuous suppression of the EI gap, consistent with an FFLO-like pairing mechanism~\cite{fulde1964superconductivity}}. In this regime, the ordering vector approaches the Fermi-surface mismatch, $q\simeq q_0=|k_{F,v}-k_{F,c}|$, consistent with experimental observations of excitonic charge-density-wave (CDW) states, where the ordering vector approximately matches the momentum separation between electron and hole pockets~\cite{x6k9-wgk9,px77-3gg1}.

{As shown in Fig.~\ref{FIG2}(c), this  finite-${\bf q}$ EI order reconstructs the momentum-space pairing landscape, and only selected portions of the Fermi surface retain nonzero EI correlations. The EI correlations are concentrated predominantly near the $y$-axis directions where the electron and hole states remain well matched, while vanishing along the orthogonal $x$-axis directions due to poor momentum overlap.} The vanishing EI correlations indicates that electrons in these momentum regions no longer participate in excitonic pairing. These states then naturally provide available phase space for SC pairing, allowing the system to further lower its free energy through the development of SC order $|\Delta_s|$ [Fig.~\ref{FIG2}(a)]. As a result, SC correlations preferentially develops in the momentum regions [Fig.~\ref{FIG2}(f)] where EI correlations are absent. {This complementary distribution correlations minimizes the mutual competition and stabilizes the coexistence phase of EI and SC orders.} 

With further electron or hole doping, the enlarged electron-hole Fermi-surface mismatch progressively suppresses the EI order, as shown in Fig.~\ref{FIG2}(a). Eventually, the EI gap vanishes, driving the system into a pure SC state. In this regime [Fig.~\ref{FIG2}(d) and (g)], the EI correlations are completely suppressed, and the SC correlations recover isotropic behavior, reflecting the absence of EI-induced rotational symmetry breaking.

\begin{figure}[htb]
  {\includegraphics[width=8.6cm]{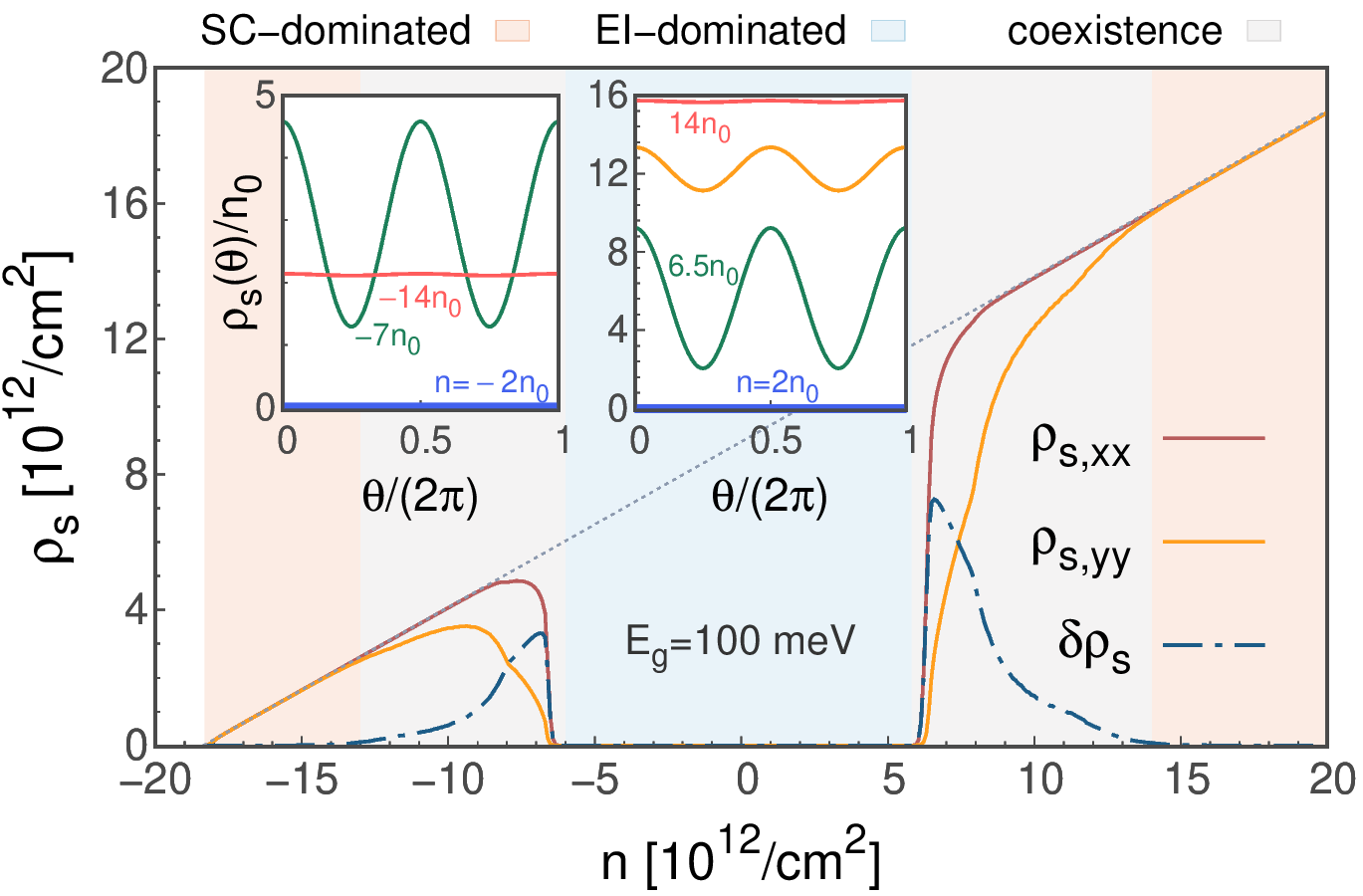}}
\caption{
Superfluid-density tensor components $\rho_{s,xx}$ and $\rho_{s,yy}$, and the nematic component
$\delta\rho_s=\rho_{s,xx}-\rho_{s,yy}$, versus net carrier density
$n=n_c-n_v$. Insets show the angular dependence of $\rho_s(\theta)$ in the EI-dominated, coexistence, and SC-dominated regimes on both the hole- and electron-doped sides. Here, $n_0=10^{12}~\mathrm{cm}^{-2}$. The dashed line denotes the conduction-band density $n_c$, where superconductivity resides; thus, in the SC-dominated regime ($|n|>14n_0$),
$\rho_{s,xx}\approx\rho_{s,yy}\approx n_c$. The phase regimes are determined from Fig.~\ref{FIG2}(a).
}
\label{FIG3}
\end{figure}

{\sl Nematic Superconductivity.---}{The normal-state band structure and electron pocket considered here preserve rotational symmetry. Therefore, the SC response in the SC-dominated regime  remains isotropic. In contrast, the finite-${\bf q}$ EI order in the coexistence regime provides a symmetry-breaking background that induces an anisotropic SC response.} As shown in Fig.~\ref{FIG3} for the calculated superfluid density tensor $\rho_{s,ij}$ (see Sec.~SIII for details), in the SC-dominated regime (e.g., $n=\pm14n_0$), $\rho_{s,ij}$ remains isotropic, consistent with the rotational symmetry of the underlying band structure.  Upon entering the EI-SC coexistence regime (e.g., $n=-7n_0$, $n=6.5n_0$, and $n=9n_0$), the superfluid density develops a pronounced $C_2$ anisotropy (insets of Fig.~\ref{FIG3}), despite the isotropic normal-state electronic structure, revealing an emergent nematic SC state driven by finite-${\bf q}$-EI order. Throughout the coexistence regime, $\rho_{s,xx}>\rho_{s,yy}$, indicating enhanced SC stiffness along the direction perpendicular to the excitonic density-wave vector $\pm{\bf q}$. The resulting nematicity is substantial, with the anisotropy ratio 
$\frac{\rho_{s,xx}-\rho_{s,yy}}{\rho_{s,xx}+\rho_{s,yy}} $
reaching $\sim70\%$ on both hole- and electron-doped sides (Fig.~SII), demonstrating a strong and experimentally accessible nematic SC response.

Physically, the superfluid density tensor can be approximately understood as $\rho_{s,ij}\sim\sum_{\bf k} k_i k_j \rho_{\rm sc}({\bf k})$. Owing to the anisotropic distribution of the momentum-resolved SC correlation $\rho_{\rm sc}({\bf k})$ in the coexistence regime [Fig.~\ref{FIG2}(f)], the contributions from different momentum directions become unequal. This directly leads to $\rho_{s,xx}>\rho_{s,yy}$ and hence to intrinsic nematic superconductivity in the EI-SC coexistence regime. 

{\sl EI-SC Phase Transitions.---}Previous studies predict a mutually exclusive relationship between EI and SC orders~\cite{PhysRevB.71.224502}, as the excitonic pairing is restricted to occur at $q=0$, leading to an isotropic distribution of EI correlations over the entire Fermi surface. Accordingly, the isotropic EI and SC pairings must compete for the same low-energy states, preventing their coexistence, and tuning the carrier density generally leads to an abrupt suppression of the EI gap beyond a critical doping level~\cite{EIdiscussion}, indicative of a first-order-like transition, after which the system enters either a normal or an SC state with vanishing EI order. In Fig.~\ref{FIG2}(a) here, owing to the unique competition via a redistribution of pairing phase space,   the SC gap $|\Delta_{s}|$ develops continuously from zero and grows as the EI gap decreases, signaling a {continuous} transfer of pairing phase space from the particle-hole channel to the particle-particle channel upon carrier doping. Such a continuous evolution of gaps is consistent with experimental observations in candidate systems exhibiting intertwined excitonic and SC correlations~\cite{Rhodes2021,x6k9-wgk9,px77-3gg1,qfqm-qp82}.

{\sl Particle-Hole Asymmetry of Doping evolution of SC Gap.---}On the electron-doped side, the SC gap increases with doping and gradually saturates at higher carrier densities, consistent with the characteristic behavior of mean-field theory~\cite{b6vp-zt8z}. On the hole-doped side, the SC gap exhibits a dome-like structure: it initially increases as the EI order is suppressed, but eventually decreases and vanishes when $n_c \rightarrow 0$, due to the progressive depletion of conduction-band electrons required for Cooper pairing. As $E_g$ is reduced, both the SC dome on the hole-doped side and the SC-EI coexistence regime shrink substantially [inset of Fig.~\ref{FIG2}(a)]. Thus, in materials with significantly small band-overlap scales, such as monolayer $\mathrm{1T'}$-\textnormal{WTe}$_2$~\cite{Song2024,PhysRevResearch.7.013224,b6vp-zt8z,Fatemi2018,Sajadi2018}, the coexistence is expected to be rather limited.

{\sl Discussion.---}Another consequence of finite-${\bf q}$ EI order is an accompanying real-space charge-density modulation with the ordering wave vector ${\bf Q}_{\rm CDW}$ (see Sec.~SV for details),
\begin{equation}
\rho({\bf r})
=
\rho_0
+
|\rho_{\bf q}|
\cos({\bf Q}_{\rm CDW}\cdot{\bf r}+\theta),
\end{equation}
where $\rho_{\bf q}
=
|\rho_{\bf q}|
e^{i\theta}
\propto
\Delta_e({\bf q})$. For multivalley systems, the  density-wave ordering vector contains both the large intervalley momentum and the residual finite pairing momentum, ${\bf Q}_{\rm CDW}
=
\Delta{\bf K}
+
{\bf q}$, where $\Delta{\bf K}$ connects the conduction- and valence-band valleys. Only the residual nesting vector ${\bf q}$ (the
remaining momentum mismatch between the electron and hole pockets after they are aligned to a
common momentum center) enters the self-consistent calculation and governs the low-energy EI instability, while $\Delta{\bf K}$ does not enter the 
calculation. 

The rotational symmetry breaking of the proposed nematic superconductivity originates from the finite-${\bf q}$ EI order, whose spontaneous symmetry breaking is imprinted onto the SC state through the coupling between the EI and SC orders.  As the excitonic pairing selects a preferred direction of $\pm{\bf q}$, SC correlations develop predominantly along the orthogonal direction, producing a $C_2$-anisotropy in superfluid transport despite an isotropic normal-state band structure of electrons. Unlike conventional scenarios in which electronic nematicity is externally induced by symmetry-breaking fields or inherited from intrinsic band anisotropy, our results suggest a continuous transition from an isotropic system to a state with a spontaneously selected direction driven by the interplay between competing orders, pointing to a potentially generic mechanism for intrinsic electronic nematicity in multi-order systems. 

The EI and SC instabilities here are treated on an equal footing within the microscopic mean-field framework, as widely employed in describing superconductivity, excitonic insulators, and other coupled electronic instabilities  (e.g., coupled spin-density-waves/itinerant-antiferromagnetic and SC instabilities in iron pnictides~\cite{PhysRevB.82.014521,PhysRevB.79.060508}). Our results demonstrates that, at moderate electron–hole Fermi-surface mismatch, the EI instability, even when significantly stronger than the SC instability, does not destroy the SC order as expected in previous studies. The coexistence emerges naturally from the redistribution of pairing phase space on Fermi surface through a self-consistent solution of two competing instabilities that minimizes the free energy. Such coexistence has been observed in candidate materials such as monolayer 1T$'$-MoTe$_2$~\cite{x6k9-wgk9,px77-3gg1}. The proposed mechanism provides a microscopic understanding of this coexistence, and calls for further experimental investigations of the nematic superconductivity through anisotropic transport measurements. The highly anisotropic SC transport response, with an anisotropy ratio of up to 70\% predicted here, represents a distinct manifestation of the finite-${\bf q}$ EI-induced momentum-space reconstruction and may yield a transport regime approaching quasi-1D  superconductivity. Doping or gating can continuously modulate EI order, driving a crossover from anisotropic to nearly isotropic SC transport and providing a route toward electrically controlled nematic superconductivity.  {These features represent characteristic predictions
of the proposed mechanism,
offering experimentally accessible signatures
of the underlying interplay between EI and SC orders}. 

In conventional CDW systems, the generic imperfect nesting of a single-band Fermi surface in two dimensions allows the CDW order to gap only part of the Fermi surface, leaving residual ungapped states that naturally support SC pairing and yields CDW-SC coexistence~\cite{PhysRevB.101.134521,PhysRevB.20.4457,PhysRevB.82.144531,Chen2016}. In EI systems, the electron and hole Fermi surfaces can be tuned to achieve nearly perfect matching, allowing the EI order to gap the entire relevant low-energy states.
 This leads to a competition between EI and SC orders for the available pairing phase space, a mechanism that is generically absent in conventional CDW systems.

{\sl Acknowledgments.---}F.Y. acknowledges valuable and insightful discussions with  Chaoxing Liu and Xiaoxiang Xi. F.Y. and J.W.L. acknowledge support from the National Key R\&D Program of China (2021YFA1401500) and the Hong Kong Research Grants Council (CRS\_HKUST603/25, C6046-24G, 16306722, 16304523, and 16311125). B.Y. acknowledges the financial support by the Penn State Materials Research Science and Engineering Center for Nanoscale Science (MRSEC) under National Science Foundation award DMR-2011839.

\bibliography{ref.bib}

\end{document}


\title{Excitonic-Superconducting Coexistence and Emergent Nematic Superconductivity Driven by Spontaneous Symmetry Breaking (Supplementary  Materials)}

\author{Fei Yang}
\thanks{These authors contributed equally to this work.}

\affiliation{Department of Physics, The Hong Kong University of Science and Technology, Hong Kong SAR, China}

\author{Ruigang Li}
\thanks{These authors contributed equally to this work.}

\affiliation{Department of Physics, National University of Singapore, Singapore 117551, Republic of Singapore}

\author{Junwei Liu}
\email{liuj@ust.hk}
\affiliation{Department of Physics, The Hong Kong University of Science and Technology, Hong Kong SAR, China}

\author{Binghai Yan}
\email{binghai.yan@psu.edu}
\affiliation{Department of Physics, The Pennsylvania State University, University Park, PA 16802, USA}
\affiliation{Center for Theory of Emergent Quantum Matter, The Pennsylvania State University, University Park, PA 16802, USA}

\maketitle 

\tableofcontents

\section{Self-consistent microscopic framework}

\subsection{Derivation of the gap equations}

In this section, we give the explicit derivation of the SC and EI gap equations.  We start from the mean-field Hamiltonian written in the Nambu basis
$\Psi_{\bf k}^{\dagger}
=(
c_{{\bf k}\uparrow}^{\dagger},
c_{-{\bf k}\downarrow},
v_{{\bf k}+{\bf q},\uparrow}^{\dagger},
v_{-{\bf k}-{\bf q},\downarrow}
)$~\cite{abrikosov2012methods},
\begin{eqnarray}
H_{\rm MF}
=
\sum_{\bf k}\Psi_{\bf k}^\dagger
{\cal H}_{\bf k}
\Psi_{\bf k}
+\frac{|\Delta_s|^2}{g_{\rm SC}}
+\frac{|\Delta_e({\bf q})|^2}{g_{\rm EI}},
\end{eqnarray}
where
\begin{equation}\label{SMMFH}
{\cal H}_{\bf k}=
\begin{pmatrix}
\xi_{c,\mathbf k} & -\Delta_s & -\Delta_e({\bf q}) & 0\\
-\Delta_s^\ast & -\xi_{c,-\mathbf k} & 0 & \Delta_e^\ast({\bf q})\\
-\Delta_e^\ast({\bf q}) & 0 & \xi_{v,{\bf k+q}} & 0\\
0 & \Delta_e({\bf q}) & 0 & -\xi_{v,{\bf -k-q}}
\end{pmatrix}.
\end{equation}
Here $\Delta_s$ denotes the intraband SC order parameter in the conduction band, while
$\Delta_e({\bf q})$ is the EI density-wave order parameter coupling the conduction band state at ${\bf k}$ to the valence band state at ${\bf k}+{\bf q}$. The system discussed here is treated within a mean-field framework~\cite{abrikosov2012methods,schrieffer1964theory,tinkham2004introduction,b6vp-zt8z,gpbp-qhp9,3q5l-46h5,xy6z-hxcv}, such that only the amplitudes of the order parameters enter the quasiparticle spectrum self-consistently. Accordingly, we take $\Delta_s=\Delta_s^*=|\Delta_s|$ and $\Delta_e=\Delta_e^*=|\Delta_e|$, analogous to the conventional BCS theory of superconductivity~\cite{PhysRev.108.1175,schrieffer1964theory,tinkham2004introduction,abrikosov2012methods,mahan2013many} and conventional EI theories~\cite{PhysRev.158.462,PhysRevLett.19.439,RevModPhys.40.755}.

In consideration of the inversion-symmetric dispersions,
$\xi_{c,-{\bf k}}=\xi_{c,{\bf k}}$ and
$\xi_{v,-{\bf k}-{\bf q}}=\xi_{v,{\bf k}+{\bf q}}$,
the characteristic equation of ${\cal H}_{\bf k}$ is even in energy,
\begin{equation}
\det(E-{\cal H}_{\bf k})
=
[E^2-E_+^2({\bf k})][E^2-E_-^2({\bf k})].
\end{equation}
Defining
\begin{equation}
A_{\bf k}
=
\xi_{c,\mathbf k}^2
+
\xi_{v,{\bf k+q}}^2
+
|\Delta_s|^2
+
2|\Delta_e|^2,
\end{equation}
one finds
\begin{equation}
E_{\pm}^2({\bf k})
=
\frac{A_{\bf k}\pm\sqrt{D_{\bf k}}}{2},
\end{equation}
with
\begin{eqnarray}
D_{\bf k}
&=&(
\xi_{c,\mathbf k}^2
-
\xi_{v,{\bf k+q}}^2
+
|\Delta_s|^2)^2+
4|\Delta_e|^2
\left[
(\xi_{c,\mathbf k}+\xi_{v,{\bf k+q}})^2
+
|\Delta_s|^2
\right].
\end{eqnarray}
Thus the quasiparticle spectrum consists of four particle-hole-related branches,
$\pm E_+({\bf k})$ and $\pm E_-({\bf k})$.

The thermodynamic potential is then~\cite{abrikosov2012methods}
\begin{equation}
\Omega_{\bf q}(|\Delta_s|,|\Delta_e|)
=
\frac{|\Delta_s|^2}{g_{\rm SC}}
+
\frac{|\Delta_e({\bf q})|^2}{g_{\rm EI}}
-
k_BT
\sum_{\bf k}
\sum_{\alpha=\pm}
\ln
\left[
2\cosh
\left(
\frac{E_\alpha({\bf k})}{2k_BT}
\right)
\right].
\end{equation}
Compared with conventional zero-momentum EI theories~\cite{PhysRev.158.462,PhysRevLett.19.439,RevModPhys.40.755,PhysRevB.71.224502,Bi2021}, the present free energy depends not only on the two gap amplitudes
$|\Delta_s|$ and $|\Delta_e|$, but also on the finite ordering wave vector ${\bf q}$.
The equilibrium state is therefore obtained by minimizing
$\Omega_{\bf q}$ with respect to all three variational parameters:
\begin{equation}
\frac{\partial \Omega_{\bf q}}{\partial |\Delta_s|}=0,
\qquad
\frac{\partial \Omega_{\bf q}}{\partial |\Delta_e|}=0,
\qquad
\mathbf q = \arg\min_{\mathbf q}\,\Omega_{\mathbf q}(|\Delta_s|,|\Delta_e|).
\end{equation}
Here, $\arg\min_{\mathbf q}\Omega_{\mathbf q}$ denotes the value of $\mathbf q$ that minimizes the thermodynamic potential $\Omega_{\mathbf q}$.\\

For a generic gap amplitude $|\Delta_i|$, where
$i=s,e$, the derivative of the quasiparticle contribution is
\begin{eqnarray}
\frac{\partial \Omega_{\bf q}}{\partial |\Delta_i|}
&=&
\frac{2|\Delta_i|}{g_i}
-
\frac{1}{2}
\sum_{{\bf k},\alpha=\pm}
\tanh
\left(
\frac{E_\alpha}{2k_BT}
\right)
\frac{\partial E_\alpha}{\partial |\Delta_i|}.
\end{eqnarray}
Since
\begin{equation}
\frac{\partial E_\alpha}{\partial |\Delta_i|}
=
\frac{1}{2E_\alpha}
\frac{\partial E_\alpha^2}{\partial |\Delta_i|},
\end{equation}
the self-consistency condition can be written as
\begin{equation}
\frac{1}{g_i}
=
\frac{1}{4|\Delta_i|}
\sum_{{\bf k},\alpha=\pm}
\tanh
\left(
\frac{E_\alpha}{2k_BT}
\right)
\frac{1}{E_\alpha}
\frac{\partial E_\alpha^2}{\partial |\Delta_i|}.
\end{equation}

For the SC gap, differentiating $E_\alpha^2$ with respect to $|\Delta_s|$ gives
\begin{equation}
\frac{\partial E_\alpha^2}{\partial |\Delta_s|}
=
|\Delta_s|
\left[
1
+
\alpha
\frac{
\xi_{c,\mathbf k}^2
-
\xi_{v,{\bf k+q}}^2
+
|\Delta_s|^2
+
2|\Delta_e|^2
}{
E_+^2-E_-^2
}
\right].
\end{equation}
This expression can be rearranged into the compact form
\begin{equation}
\frac{1}{g_{\rm SC}}
=
-
\sum_{{\bf k},\alpha=\pm}
\frac{
\alpha
\left(
E_\alpha^2-\xi_{v,{\bf k+q}}^2
\right)
}{
2E_\alpha
\left(
E_+^2-E_-^2
\right)
}
\tanh
\left(
\frac{E_\alpha}{2k_BT}
\right).
\label{SCGE}
\end{equation}
Equation~(\ref{SCGE}) shows that the SC gap is renormalized by the \emph{anisotropic} EI hybridization through both the quasiparticle energies $E_\alpha$ and the spectral-weight factor
$(E_\alpha^2-\xi_{v,{\bf k+q}}^2)/(E_+^2-E_-^2)$.

Similarly, differentiating $E_\alpha^2$ with respect to $|\Delta_e|$ yields
\begin{equation}
\frac{\partial E_\alpha^2}{\partial |\Delta_e|}
=
2|\Delta_e|
\left[
1
+
\alpha
\frac{
(\xi_{c,\mathbf k}+\xi_{v,{\bf k+q}})^2
+
|\Delta_s|^2
}{
E_+^2-E_-^2
}
\right].
\end{equation}
After rearrangement, the EI gap equation becomes
\begin{equation}
\frac{1}{g_{\rm EI}}
=
-
\sum_{{\bf k},\alpha=\pm}
\frac{
\alpha
\left(
E_\alpha^2
-
|\Delta_e|^2
+
\xi_{c,\mathbf k}\xi_{v,{\bf k+q}}
\right)
}{
2E_\alpha
\left(
E_+^2-E_-^2
\right)
}
\tanh
\left(
\frac{E_\alpha}{2k_BT}
\right).
\label{EIGE}
\end{equation}
Equations~(\ref{SCGE}) and~(\ref{EIGE}), supplemented by the condition
\begin{equation}
\mathbf q = \arg\min_{\mathbf q}\,\Omega_{\mathbf q}(|\Delta_s|,|\Delta_e|),
\end{equation}
form a closed set of mean-field equations describing superconductivity and a finite-momentum EI density-wave order.\\

In general, the interband Coulomb interaction can be separated into
direct and exchange contributions,
\begin{equation}
H_{cv}
=
H_{cv}^{\mathrm{dir}}
+
H_{cv}^{\mathrm{ex}},
\end{equation}
where the conduction-band electrons are centered around the $K$ valley,
while the valence-band electrons are centered around the inequivalent
$K'$ valley. The direct Coulomb interaction  is given by
\begin{equation}
H_{cv}^{\mathrm{dir}}
=
\sum_{\mathbf{k},\mathbf{k}',\mathbf{q}}
V_{cv}^{\mathrm{dir}}(\mathbf q)\,
c^{\dagger}_{\mathbf K+\mathbf{k}+\mathbf q}
v^{\dagger}_{\mathbf K'+\mathbf{k}'-\mathbf q}
v_{\mathbf K'+\mathbf{k}'}
c_{\mathbf K+\mathbf{k}}=
-
\sum_{\mathbf{k},\mathbf{k}',\mathbf{q}}
V_{cv}^{\mathrm{dir}}(\mathbf q)\,
c^{\dagger}_{\mathbf K+\mathbf{k}+\mathbf q}
v^{\dagger}_{\mathbf K'+\mathbf{k}'-\mathbf q}
c_{\mathbf K+\mathbf{k}}
v_{\mathbf K'+\mathbf{k}'}
.
\end{equation}

The exchange interaction  takes the form
\begin{equation}
H_{cv}^{\mathrm{ex}}
=
-
\sum_{\mathbf{k},\mathbf{k}',\mathbf{q}}
V_{cv}^{\mathrm{ex}}
(\Delta\mathbf K+\mathbf{k}'-\mathbf{k}-\mathbf q)
\,
c^{\dagger}_{\mathbf K+\mathbf{k}+\mathbf q}
v^{\dagger}_{\mathbf K'+\mathbf{k}'-\mathbf q}
c_{\mathbf K+\mathbf{k}}
v_{\mathbf K'+\mathbf{k}'},
\end{equation}
where
\begin{equation}
\Delta\mathbf K=\mathbf K'-\mathbf K
\end{equation}
is the intervalley momentum transfer connecting the two valleys.

Physically, the direct interaction mainly involves small-momentum-transfer
scattering processes and therefore provides the dominant attractive
interaction responsible for exciton formation and EI condensation,
as commonly discussed in the literature~\cite{PhysRev.158.462,PhysRevLett.19.439,RevModPhys.40.755,Bi2021,PhysRevB.71.224502}.
By contrast, the exchange interaction necessarily involves a large
momentum transfer of order $|\mathbf K-\mathbf K'|$, making it
considerably weaker in most EI systems.
Consequently, the exchange term usually acts only as a perturbative
correction. 

\subsection{Exact diagonalization of the mean-field Hamiltonian}
\label{sdsd}

Here we present the exact diagonalization of the mean-field  Hamiltonian. 
For simplicity, we choose a global gauge such that both order parameters are real: $\Delta_{\rm s},\Delta_{\rm e}\in\mathbb R$. For compact notation, we define
\begin{equation}
\xi_c
\equiv
\xi_{c,{\bf k}},
\qquad
\xi_v
\equiv
\xi_{v,{\bf k}+{\bf q}},
\qquad
\Delta_s
\equiv
|\Delta_{\rm s}|,
\qquad
\Delta_e
\equiv
|\Delta_{\rm e}|.
\end{equation}
Then, the mean-field Hamiltonian (\ref{SMMFH}) is written as
\begin{equation}
\mathcal H_{\bf k}
=
\begin{pmatrix}
\xi_c
&
-\Delta_s
&
-\Delta_e
&
0
\\
-\Delta_s
&
-\xi_c
&
0
&
\Delta_e
\\
-\Delta_e
&
0
&
\xi_v
&
0
\\
0
&
\Delta_e
&
0
&
-\xi_v
\end{pmatrix},
\label{BdG_appendix}
\end{equation}
which is Hermitian: $
\mathcal H_{\bf k}^\dagger
=
\mathcal H_{\bf k}$. The characteristic polynomial is
\begin{equation}
\det
\left[
\epsilon
-
\mathcal H_{\bf k}
\right]
=
(\epsilon^2-E_+^2)
(\epsilon^2-E_-^2),
\end{equation}
where
\begin{equation}
E_{\pm}^2
=
\frac{
A_{\bf k}
\pm
\sqrt{D_{\bf k}}
}{2},
\end{equation}
with
\begin{equation}
A_{\bf k}
=
\xi_c^2
+
\xi_v^2
+
\Delta_s^2
+
2\Delta_e^2, \quad 
D_{\bf k}
=
\left(
\xi_c^2
-
\xi_v^2
+
\Delta_s^2
\right)^2
+
4\Delta_e^2
\left[
(\xi_c+\xi_v)^2
+
\Delta_s^2
\right].
\end{equation}

The quasiparticle eigenvalues are therefore
\begin{equation}
\epsilon_1
=
E_+,
\qquad
\epsilon_2
=
-E_+,
\qquad
\epsilon_3
=
E_-,
\qquad
\epsilon_4
=
-E_-.
\end{equation}

The spectrum exhibits the usual BdG particle-hole symmetry~\cite{PhysRev.108.1175,schrieffer1964theory,tinkham2004introduction,abrikosov2012methods,mahan2013many},
\begin{equation}
\epsilon
\leftrightarrow
-\epsilon,
\end{equation}
leading to the paired structure
\((E_+,-E_+)\) and \((E_-,-E_-)\).

The corresponding normalized eigenvectors can be written compactly as
\begin{equation}
u_\lambda
=
\frac{1}{\mathcal N_\lambda}
\begin{pmatrix}
\Delta_s(\xi_v^2-\epsilon_\lambda^2)
\\[0.8em]
X_\lambda(\xi_v+\epsilon_\lambda)
\\[0.8em]
\Delta_e\Delta_s(\xi_v+\epsilon_\lambda)
\\[0.8em]
\Delta_e X_\lambda
\end{pmatrix},
\qquad
\lambda=1,2,3,4,
\end{equation}
where
\begin{equation}
X_\lambda
=
(\xi_c-\epsilon_\lambda)
(\xi_v-\epsilon_\lambda)
-
\Delta_e^2,
\end{equation}
and the normalization factor 
\begin{eqnarray}
\mathcal N_\lambda=
\Big[
\Delta_s^2
(\xi_v^2-\epsilon_\lambda^2)^2
+
X_\lambda^2
(\xi_v+\epsilon_\lambda)^2
+
\Delta_e^2\Delta_s^2
(\xi_v+\epsilon_\lambda)^2
+
\Delta_e^2X_\lambda^2
\Big]^{1/2}.
\end{eqnarray}

The diagonalizing unitary matrix is then constructed as $U_{\bf k}
=
\left(
u_1,
u_2,
u_3,
u_4
\right)$, which takes the form:
\begin{equation}
U_{\bf k}
=
\begin{pmatrix}
\dfrac{\Delta_s(\xi_v^2-\epsilon_1^2)}{\mathcal N_1}
&
\dfrac{\Delta_s(\xi_v^2-\epsilon_2^2)}{\mathcal N_2}
&
\dfrac{\Delta_s(\xi_v^2-\epsilon_3^2)}{\mathcal N_3}
&
\dfrac{\Delta_s(\xi_v^2-\epsilon_4^2)}{\mathcal N_4}
\\[1.2em]
\dfrac{X_1(\xi_v+\epsilon_1)}{\mathcal N_1}
&
\dfrac{X_2(\xi_v+\epsilon_2)}{\mathcal N_2}
&
\dfrac{X_3(\xi_v+\epsilon_3)}{\mathcal N_3}
&
\dfrac{X_4(\xi_v+\epsilon_4)}{\mathcal N_4}
\\[1.2em]
\dfrac{\Delta_e\Delta_s(\xi_v+\epsilon_1)}{\mathcal N_1}
&
\dfrac{\Delta_e\Delta_s(\xi_v+\epsilon_2)}{\mathcal N_2}
&
\dfrac{\Delta_e\Delta_s(\xi_v+\epsilon_3)}{\mathcal N_3}
&
\dfrac{\Delta_e\Delta_s(\xi_v+\epsilon_4)}{\mathcal N_4}
\\[1.2em]
\dfrac{\Delta_e X_1}{\mathcal N_1}
&
\dfrac{\Delta_e X_2}{\mathcal N_2}
&
\dfrac{\Delta_e X_3}{\mathcal N_3}
&
\dfrac{\Delta_e X_4}{\mathcal N_4}
\end{pmatrix}.
\label{Uk_explicit}
\end{equation}

One may verify directly that
\begin{equation}
\mathcal H_{\bf k}
u_\lambda
=
\epsilon_\lambda
u_\lambda .
\end{equation}

Since \(\mathcal H_{\bf k}\) is Hermitian, eigenvectors corresponding to distinct eigenvalues are mutually orthogonal:
\begin{equation}
u_\lambda^\dagger
u_{\lambda'}
=
\delta_{\lambda\lambda'}.
\end{equation}
Therefore,
\begin{equation}
U_{\bf k}^\dagger
U_{\bf k}
=
I.
\end{equation}

The Hamiltonian is thus diagonalized as
\begin{equation}
U_{\bf k}^{\dagger}
\mathcal H_{\bf k}
U_{\bf k}
=
{\rm diag}
\left(
\epsilon_1,
\epsilon_2,
\epsilon_3,
\epsilon_4
\right)
=
{\rm diag}
\left(
E_+,
-E_+,
E_-,
-E_-
\right).
\end{equation}

\subsection{Time-Reversal Symmetry of the LO-Like EI State}
\label{secLO}

In the context of the present study, we have considered the more stable LO-like EI state, where the low-energy momentum space is partitioned into two complementary sectors. 
Electron-hole pairing with center-of-mass momentum $+{\bf q}$ occurs only within one half of momentum space that is energetically favorable for finite-momentum pairing, while the opposite half is paired with $-{\bf q}$. 
Accordingly, the EI order parameter can be written schematically as
\begin{equation}
\Delta_{+}
\sim
\sum_{{\bf k}\in R,s}
\langle
v^\dagger_{{\bf k}+{\bf q},s}
c_{{\bf k},s}
\rangle,
\end{equation}
and
\begin{equation}
\Delta_{-}
\sim
\sum_{{\bf k}\in L,s}
\langle
v^\dagger_{{\bf k}-{\bf q},s}
c_{{\bf k},s}
\rangle,
\end{equation}
where the two momentum sectors satisfy
\begin{equation}
R\cup L=\mathcal P,
\qquad
R\cap L=\emptyset.
\end{equation}

Under time-reversal symmetry, momentum transforms as
\begin{equation}
{\bf k}\rightarrow-{\bf k},
\qquad
{\bf q}\rightarrow-{\bf q}.
\end{equation}
Therefore, a single finite-${\bf q}$ EI component is not invariant under time reversal by itself, since the $+{\bf q}$ pairing sector is mapped to the corresponding $-{\bf q}$ sector. 
However, in the LO-like construction considered here, the two momentum sectors $R$ and $L$ are related by time reversal. 
Consequently, the $+{\bf q}$ EI component defined in $R$ is transformed into the $-{\bf q}$ component defined in $L$, and vice versa. 
The two sectors therefore form a time-reversal pair, such that the combined LO-like EI state preserves the overall time-reversal symmetry. This construction is the particle-hole analogue of the conventional LO SC state~\cite{larkin1965zh}, where opposite finite pairing momenta coexist to restore the symmetry broken by an individual Fulde-Ferrell-type component. 
At the same time, the finite ordering vector $\pm{\bf q}$ still spontaneously selects a preferred spatial direction, thereby breaking rotational symmetry and generating anisotropic EI correlations in momentum space. In particular, since neither the present system nor the related materials contain an explicit source of time-reversal-symmetry breaking, the LO-like state is energetically favored over a single-${\bf q}$ EI state, which would correspond to an FF-type construction~\cite{fulde1964superconductivity} and spontaneously break time-reversal symmetry by selecting only one ordering momentum.

For the multivalley system, such as monolayer 1T$'$-MoTe$_2$, the low-energy band structure contains two conduction-band valleys centered at $\pm{\bf K}_c$, while the valence-band maximum is located near $\Gamma$. 
The momentum variable ${\bf k}$ used in our effective theory should still be understood as spanning only one representative half of the momentum space within a given valley. 
Its time-reversal-related counterpart is then naturally mapped to the complementary half-momentum sector in the opposite valley. 
A finite-momentum EI order connects valence-band states to conduction-band states through the ordering vectors
\begin{equation}
{\bf Q}_{\pm}
=
\pm({\bf K}_c+{\bf q}).
\end{equation}
Then, it is more convenient to combine the right-moving sector of the $+{\bf K}_c$ valley and the left-moving sector of the $-{\bf K}_c$ valley into a new reconstructed conduction band, $\tilde c_{{\bf k}s}$. 
These two sectors are related by time-reversal symmetry and therefore naturally form a time-reversal-symmetric finite-momentum EI sector. 
The reconstructed band $\tilde c_{{\bf k}s}$ directly participates in the EI hybridization with the valence band. 
Equivalently, one part of $\tilde c_{{\bf k}s}$ is associated with the $+{\bf q}$ EI component, while the other part is associated with the $-{\bf q}$ component. 
Consequently, after valley folding, the resulting effective theory becomes similar to the LO-type finite-${\bf q}$ EI model used in the main text, where the $\pm{\bf q}$ components are connected by time-reversal symmetry.\\

\subsection{Transition from nearly $q=0$ EI state to the finite-${\bf q}$ EI state}

\begin{figure}[h]
    \centering
    \includegraphics[width=0.8\linewidth]{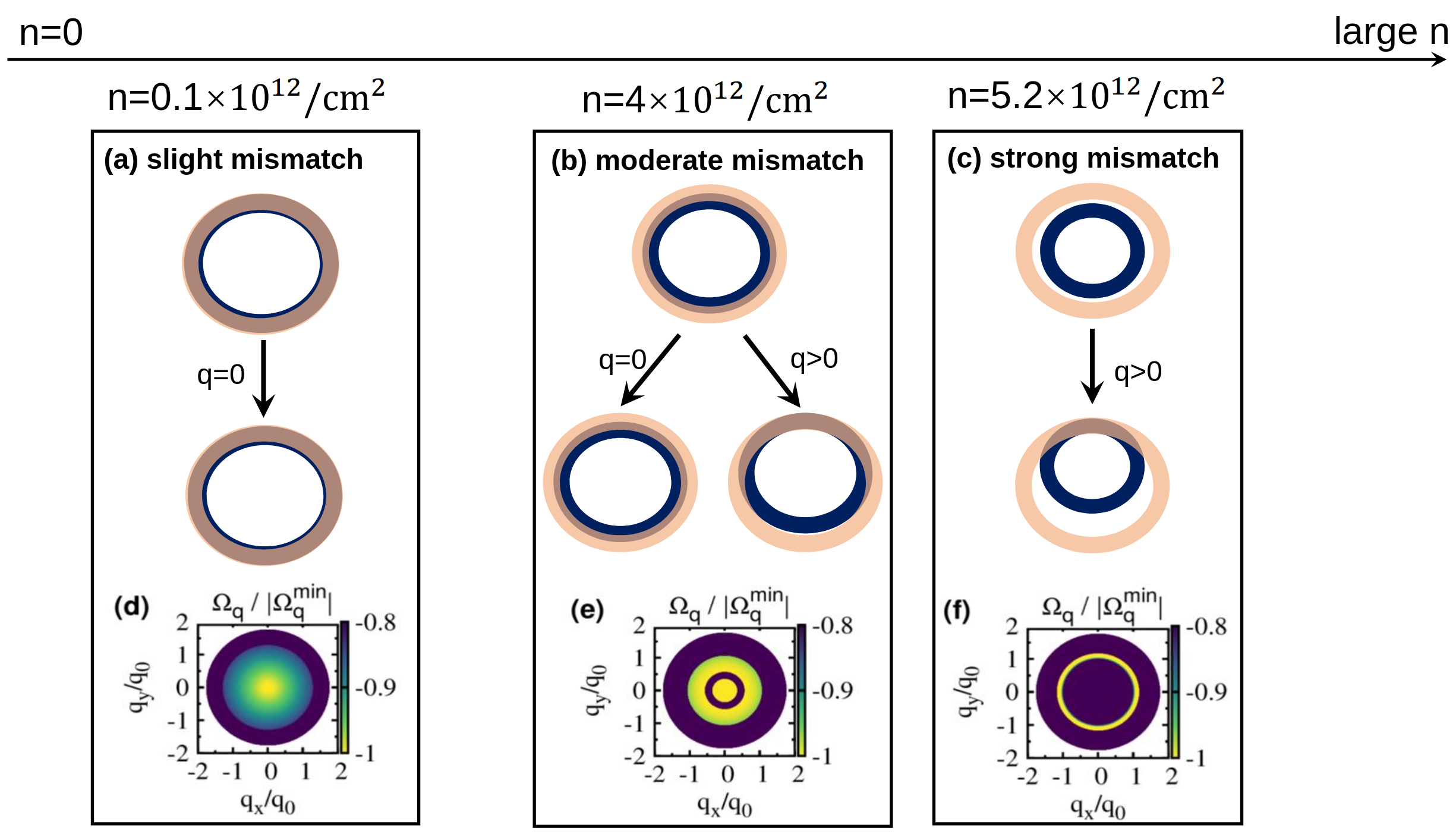}
\caption{{
(a)-(c) Schematic illustration of the electron and hole Fermi surfaces (with the effective energy windows) at three representative doping levels. We show the single-$q$ case here for simplicity of illustration. The orange and blue filled annular regions denote the effective low-energy windows around the electron and hole Fermi surfaces, respectively. The overlap between the electron and shifted-hole Fermi surface identifies the region contributing predominantly to EI pairing and determines the available phase space for EI pairing. (d)-(f) Corresponding free-energy landscapes $\Omega_{\bf q}$ as a function of the ordering wave vector ${\bf q}$, obtained from the full self-consistent calculations in the present work. Here, $\Omega_{\bf q}$ is normalized by the absolute value of its global minimum, $|\Omega_{\bf q}^{\rm min}|$. $E_g=100~$meV.}}
\label{figR1}
\end{figure}

{The microscopic mechanism for the transition from the nearly uniform $q=0$ EI state to the finite-$q$ EI state is governed by the geometry of electron-hole Fermi-surface matching, which is analogous to the well-known FFLO mechanism in superconductors~\cite{yang2018fulde,larkin1965zh,fulde1964superconductivity,Dong_2013,PhysRevA.89.013607,PhysRevLett.89.227002,PhysRevB.76.014522,PhysRevB.75.064511,PhysRevLett.114.110401,rqp1-jtcb}. To elucidate this mechanism more clearly, we provide representative electron and hole Fermi surfaces at three characteristic doping levels, as shown in Fig.~\ref{figR1}(a)-(c). For illustration, we indicate effective energy windows representing the relevant low-energy regions around the electron and hole Fermi surfaces. The overlap between these two energy windows identifies the region that contributes predominantly to EI pairing and determines the available pairing phase space. A larger overlap generally leads to a stronger EI instability, a larger self-consistent EI order parameter, and a lower free energy.}

{When the electron and hole Fermi surfaces are perfectly matched or only weakly mismatched [see Fig.~\ref{figR1}(a)], the $q=0$ EI pairing provides the largest overlap between the electron and hole pairing windows. Consequently, the free-energy landscape $\Omega_{\bf q}$ exhibits only a single extremum at ${\bf q}=0$ [see Fig.~\ref{figR1}(d)].}

{When the electron-hole Fermi surfaces become moderately mismatched upon increasing the electron or hole doping [see Fig.~\ref{figR1}(b)], both the $q=0$ and finite-$q$ pairing configurations can still generate a finite overlap between the electron and hole pairing windows. 
Accordingly, the free-energy landscape $\Omega_{\bf q}$ develops a double-well structure with two competing local minima [see Fig.~\ref{figR1}(e)].  As the electron-hole Fermi-surface mismatch increases, the global minimum switches from the $q=0$ branch to the finite-$q$ branch, thereby driving a first-order transition from the nearly uniform $q=0$ EI state to the finite-$q$ EI state. Therefore, the abrupt onset of finite-$q$ in Fig.~2(a) originates from the switching of the global free-energy minimum between these two competing states.}

{With further doping, the Fermi-surface mismatch becomes sufficiently large that the $q=0$ pairing shells no longer overlap [see Fig.~\ref{figR1}(c)].
In this regime, only the finite-$q$ solution provides electron-hole pairing phase space and remains as the stable minimum of the free energy see Fig.~\ref{figR1}(f).}\\

\begin{figure}[h]
    \centering
    \includegraphics[width=0.79\linewidth]{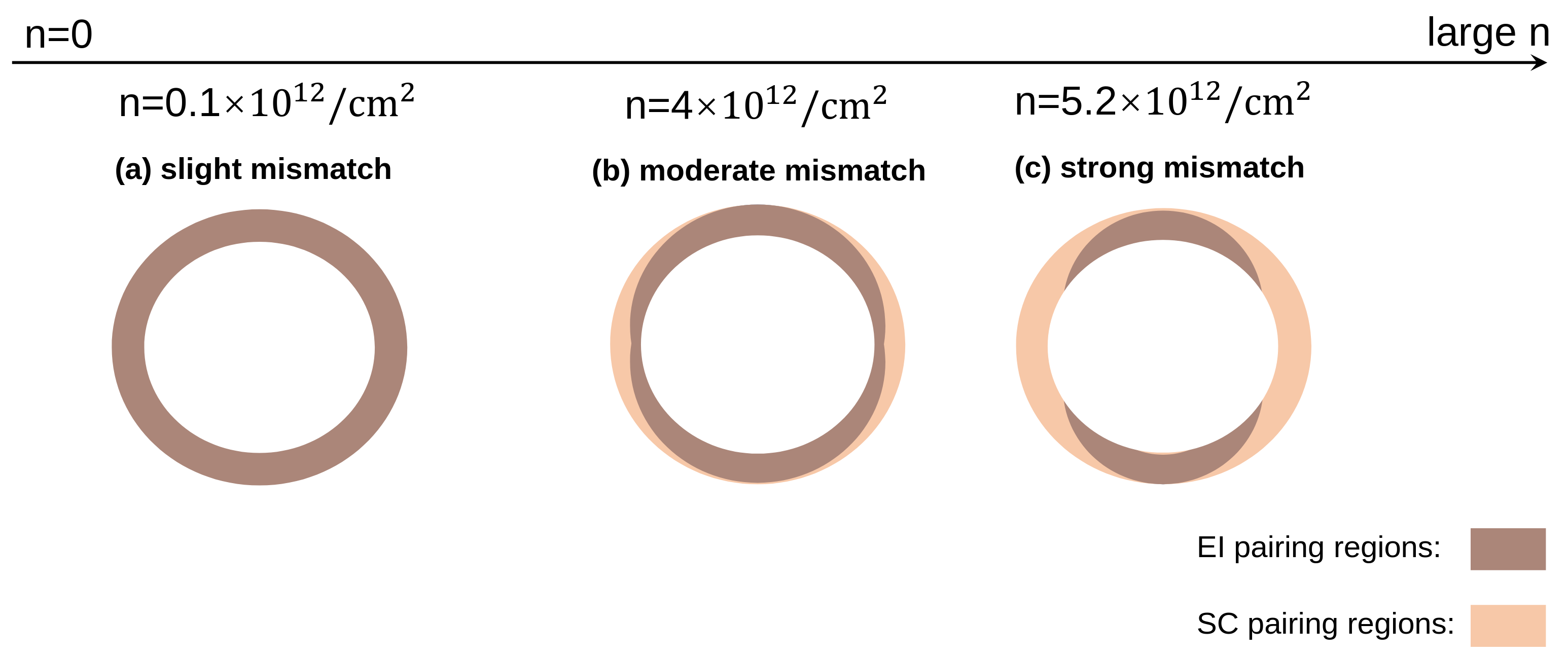}
\caption{{
Schematic illustration of the electron Fermi surfaces (with the effective energy windows)  at representative doping cases. The shaded regions here correspond to the overlap regions between the electron and shifted-hole Fermi surfaces shown in Fig.~\ref{figR1}(a)-(c). Here, the illustration is generalized to the symmetric LO-like excitonic state. The shaded regions are associated with the EI correlation, whereas the remaining momentum regions of the electron Fermi surface provide the complementary phase space for SC correlation.}}
\label{figR3}
\end{figure}

{The analysis presented in Fig.~\ref{figR1} illustrates the optimal phase-space matching for EI pairing. For further illustration, Fig.~\ref{figR3} shows the electron Fermi surface and highlights the overlapping regions with the shifted-hole Fermi surface that remain available for EI pairing. These overlapping regions, originating from the electron-hole Fermi-surface matching illustrated in Fig.~\ref{figR1} and extended to the symmetric LO-like excitonic state, correspond to the momentum sectors with finite EI correlations. In contrast, the remaining regions of the electron Fermi surface provide the complementary phase space for SC correlations. These results demonstrate the evolution of the Fermi-surface matching condition and the associated redistribution of the pairing phase space.}

\section{EI and SC correlations}

\subsection{Derivation of correlations}

To further characterize the coexistence regime between superconductivity and the finite-momentum EI order, it is useful to analyze the corresponding correlation functions and susceptibilities.
Within the present mean-field framework, the SC and EI  orders originate from different fermionic bilinears~\cite{abrikosov2012methods,mahan2013many},
\begin{equation}
\Delta_s
=
g_{\rm SC}
\sum_{\bf k}
\langle
c_{-{\bf k}\downarrow}
c_{{\bf k}\uparrow}
\rangle,
\end{equation}
and
\begin{equation}
\Delta_e({\bf q})
=
g_{\rm EI}
\sum_{\bf k}
\langle
v_{{\bf k}+{\bf q},\sigma}^{\dagger}
c_{{\bf k},\sigma}
\rangle,
\end{equation}
corresponding respectively to Cooper pairing and interband particle-hole condensation. Here, $\rho_{\rm EI}({\bf k})$ and $\rho_{\rm SC}({\bf k})$ denote the EI and SC correlations in momentum space.

The interplay between these two competing or coexisting orders can be quantified through the anomalous Green's functions in Nambu space~\cite{abrikosov2012methods,mahan2013many}.
Defining the Matsubara Green's function
\begin{equation}
{\cal G}({\bf k},i\omega_n)
=
(i\omega_n-{\cal H}_{\bf k})^{-1},
\end{equation}
the SC anomalous propagator is given by
\begin{equation}
F_s({\bf k},i\omega_n)
=
\langle
c_{{\bf k}\uparrow}
c_{-{\bf k}\downarrow}
\rangle
=
{\cal G}_{12}({\bf k},i\omega_n),
\end{equation}
while the EI anomalous propagator is
\begin{equation}
F_e({\bf k},i\omega_n)
=
\langle
v_{{\bf k}+{\bf q},\uparrow}^{\dagger}
c_{{\bf k}\uparrow}
\rangle
=
{\cal G}_{13}({\bf k},i\omega_n).
\end{equation}

After matrix inversion, one obtains
\begin{equation}
F_s({\bf k},i\omega_n)
=
\frac{
|\Delta_s|
\left[
(i\omega_n)^2
-
\xi_{v,{\bf k+q}}^2
\right]
}{
[(i\omega_n)^2-E_+^2]
[(i\omega_n)^2-E_-^2]
},\qquad
F_e({\bf k},i\omega_n)
=
\frac{
|\Delta_e|
\left[
(i\omega_n)^2
+
\xi_{c,\mathbf k}\xi_{v,{\bf k+q}}
-
|\Delta_e|^2
\right]
}{
[(i\omega_n)^2-E_+^2]
[(i\omega_n)^2-E_-^2]
}.
\end{equation}
These expressions explicitly demonstrate that the SC  and EI sectors are strongly hybridized through the common quasiparticle poles $E_\pm$. The equal-time order-parameter correlations can then be obtained through Matsubara summation:
\begin{equation}
\rho_{\rm SC}({\bf k})
=T
\sum_n
F_s({\bf k},i\omega_n),\quad\quad
\rho_{\rm EI}({\bf k})
=T
\sum_n
F_e({\bf k},i\omega_n).
\end{equation}
Carrying out the Matsubara frequency summation yields
\begin{eqnarray}
\rho_{\rm SC}({\bf k})
&=&
-
\sum_{\alpha=\pm}
\frac{
\alpha
|\Delta_s|
(E_\alpha^2-\xi_{v,{\bf k+q}}^2)
}{
2E_\alpha(E_+^2-E_-^2)
}
\tanh
\frac{E_\alpha}{2k_BT},
\end{eqnarray}
and
\begin{eqnarray}
\rho_{\rm EI}({\bf k})
&=&
-
\sum_{\alpha=\pm}
\frac{
\alpha
|\Delta_e|
(E_\alpha^2-|\Delta_e|^2+\xi_{c,\mathbf k}\xi_{v,{\bf k+q}})
}{
2E_\alpha(E_+^2-E_-^2)
}
\tanh
\frac{E_\alpha}{2k_BT}.
\end{eqnarray}
The quantities $\rho_{\rm SC}({\bf k})$ and $\rho_{\rm EI}({\bf k})$ measure the momentum-resolved participation of electronic states in SC pairing and EI particle-hole pairing, respectively, thereby identifying the regions in momentum space that contribute most strongly to the formation of the corresponding SC and EI condensates. A vanishing correlation function indicates that the corresponding electronic state does not participate in the associated pairing channel~\cite{yang2018fulde,larkin1965zh,fulde1964superconductivity,Dong_2013,PhysRevA.89.013607,PhysRevLett.89.227002,PhysRevB.76.014522,PhysRevB.75.064511,PhysRevLett.114.110401,rqp1-jtcb,PhysRevB.98.094507,PhysRevB.106.144509}.

It should be emphasized that the momentum variable ${\bf k}$ used here labels the conduction-band electron states. Therefore, the correlations $\rho_{\rm SC}({\bf k})$ and $\rho_{\rm EI}({\bf k})$ discussed in this section should be understood as conduction-electron correlations, characterizing how conduction electrons participate in the SC and EI  channels within the projected low-energy description.

\subsection{Evolution of correlations}

\begin{figure}[h]
    \centering    \includegraphics[width=1.00\linewidth]{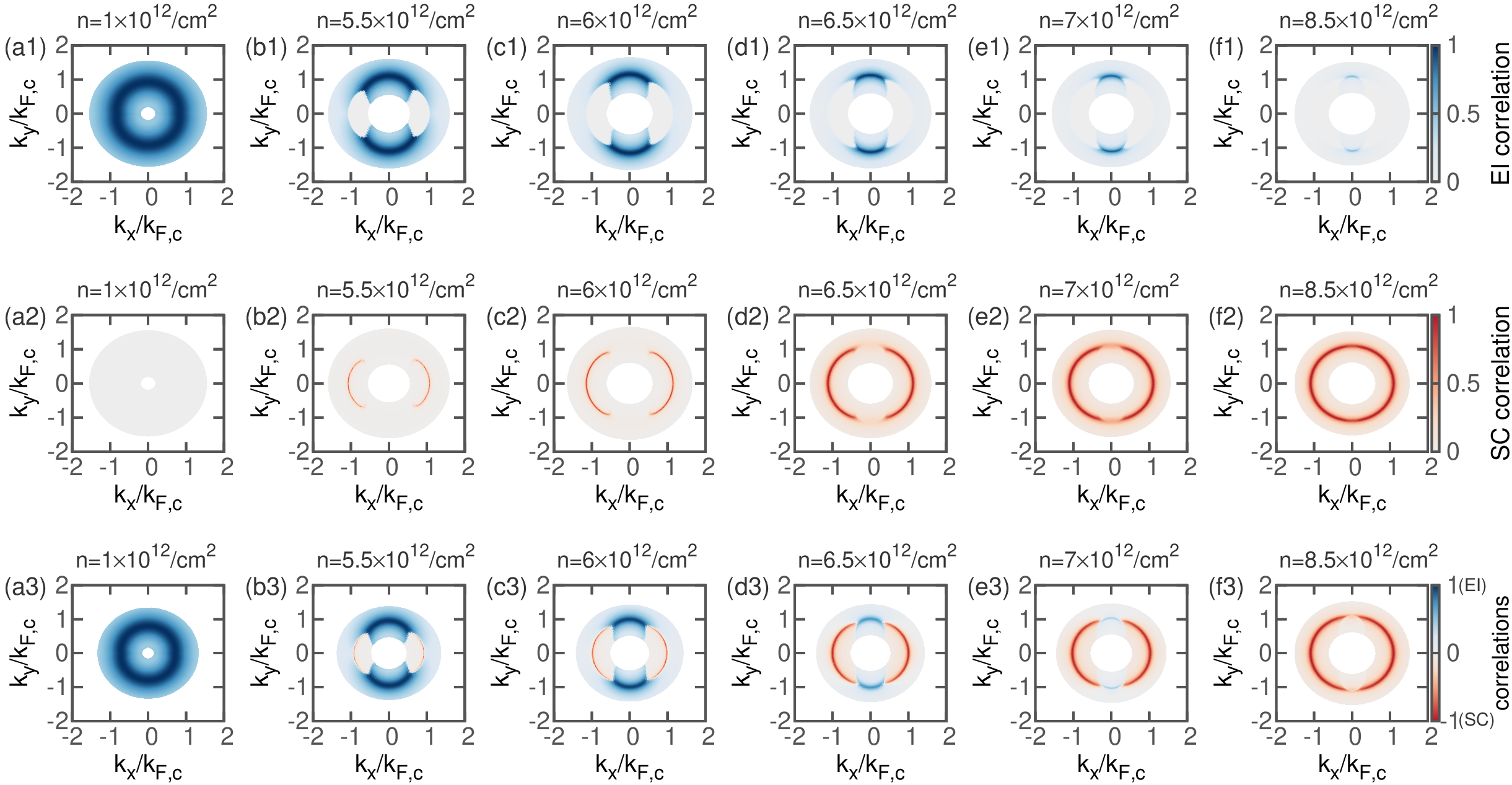}
\caption{{Momentum-resolved EI [(a1)-(f1)] and SC [(a2)-(f2)] correlations, $\rho_{\rm EI}({\bf k})=\langle v_{{\bf k}+{\bf q},s}^{\dagger}c_{{\bf k},s}\rangle$ and $\rho_{\rm SC}({\bf k})=\langle c_{{\bf k}\uparrow}c_{-{\bf k}\downarrow}\rangle$, at different carrier doping levels obtained from the self-consistent calculations. The results illustrate the evolution and redistribution of the ordering phase space between the EI and SC correlations across the EI, coexistence, and SC-dominated regimes. To further visualize the EI-SC coexistence and competition through the complementary redistribution of phase space on the Fermi surface, panels (a3)-(f3) show $\delta\rho({\bf k})=\rho_{\rm EI}({\bf k})-\rho_{\rm SC}({\bf k})$, where positive and negative values indicate momentum regions with finite EI and SC correlations, respectively, thereby revealing their momentum-space competition and coexistence. $E_g=100~$meV.}}
\label{figR2}
\end{figure}

{In this part, we present additional self-consistent calculations over a broader range of carrier densities to explicitly track the momentum-resolved EI correlation,  $\rho_{\rm EI}({\bf k})
=\langle
v_{{\bf k}+{\bf q},s}^{\dagger}
c_{{\bf k},s}\rangle$, and the SC correlation, $
\rho_{\rm SC}({\bf k})
=\langle
c_{{\bf k}\uparrow}
c_{-{\bf k}\downarrow}\rangle$. The corresponding results are summarized in Fig.~\ref{figR2}.}

{As seen from the figure, in the EI-dominated regime [Fig.~\ref{figR2}(a1) and (a2)], the momentum-resolved EI correlation occupies nearly the entire available phase space on the Fermi surface, leaving essentially no complementary momentum sectors available for the SC correlation.}

{Upon doping into the coexistence regime [Fig.~\ref{figR2}(b1)-(e1) and (b2)-(e2)] where the finite-$q$ EI order develops, the momentum-resolved EI correlation occupies only selected regions of the Fermi surface and therefore becomes strongly anisotropic, while complementary momentum sectors are released for SC Cooper pairing. As a result, the SC correlation first develops in these complementary regions and coexists self-consistently with the EI correlation.  With increasing doping within the coexistence regime, the momentum-space region occupied by the EI correlation gradually shrinks on the Fermi surface, whereas the complementary region supporting the SC correlation continuously expands. Eventually, upon  doping into the SC-dominated regime [e.g., see Fig.~\ref{figR2}(f1) and (f2)], the SC correlation occupies nearly the entire Fermi surface, while the EI correlation becomes nearly negligible. Consequently, the available pairing phase space is almost completely dominated by the SC correlation.}

{This evolution directly demonstrates how the self-consistent evolution of the finite ordering wave vector, driven by the increasing electron-hole Fermi-surface mismatch via doping, continuously redistributes the available pairing phase space between the EI and SC correlations.}

\section{Superfluid-density tensor}

In this section, we derive the superfluid density tensor by imposing a uniform SC phase twist~\cite{schrieffer1964theory},
\begin{equation}
\Delta_{\rm s}({\bf r})
=
\Delta_{\rm s} e^{2i{\bf Q}_{\rm s}\cdot{\bf r}} .
\end{equation}
The physical SC phase gradient is therefore~\cite{nambu2009nobel,nambu1960quasi,PhysRevB.100.104513,PhysRevB.98.094507,ambegaokar61electromagnetic,littlewood81gauge,schrieffer1964theory}
\begin{equation}
\nabla\theta_{\rm s}/2={\bf Q}_{\rm s}.
\end{equation}
After a gauge transformation which removes the spatial phase from the pairing term~\cite{nambu2009nobel,nambu1960quasi,PhysRevB.100.104513,PhysRevB.98.094507,ambegaokar61electromagnetic,littlewood81gauge,schrieffer1964theory}, the phase twist is transferred to the conduction-band fermions. The paired conduction electrons carry momenta
\begin{equation}
{\bf k}+{\bf Q}_{\rm s},
\qquad
-{\bf k}+{\bf Q}_{\rm s}.
\end{equation}

For the parabolic conduction-band dispersion $\xi_{c,{\bf k}}
=
-\frac{E_g}{2}
+
\frac{\hbar^2 k^2}{2m_c}
-\mu$,  we have
\begin{equation}
\xi_{c,{\bf k}+{\bf Q}_{\rm s}}
=
\xi_{c,{\bf k}}
+
{\bf v}_{c,{\bf k}}\cdot{\bf Q}_{\rm s}
+
\frac{\hbar^2 Q_{\rm s}^2}{2m_c}, \quad\quad 
\xi_{c,-{\bf k}+{\bf Q}_{\rm s}}
=
\xi_{c,{\bf k}}
-
{\bf v}_{c,{\bf k}}\cdot{\bf Q}_{\rm s}
+
\frac{\hbar^2 Q_{\rm s}^2}{2m_c},
\end{equation}
where $
{\bf v}_{c,{\bf k}}
=
\frac{\partial \xi_{c,{\bf k}}}{\partial {\bf k}}
=
\frac{\hbar^2{\bf k}}{m_c}$.  The quadratic contribution ${\hbar^2Q_{\rm s}^2}/({2m_c})$ 
generates the diamagnetic response, while the linear Doppler shift $
{\bf v}_{c,{\bf k}}\cdot{\bf Q}_{\rm s}$ 
produces the paramagnetic current response via second-order effect~\cite{schrieffer1964theory,abrikosov2012methods,PhysRevB.106.144509}. It is therefore convenient to separate the free-energy response into diamagnetic and paramagnetic contributions.

In the Nambu basis $
\Psi_{\bf k}^{\dagger}
=(
c_{{\bf k}+{\bf Q}_{\rm s},\uparrow}^{\dagger},
c_{-{\bf k}+{\bf Q}_{\rm s},\downarrow},
v_{{\bf k}+{\bf q},\uparrow}^{\dagger},
v_{-{\bf k}-{\bf q},\downarrow}
)$, the phase-twisted mean-field Hamiltonian is written as~\cite{abrikosov2012methods}
\begin{equation}
H_{\rm MF}({\bf Q}_{\rm s})
=
\sum_{\bf k}
\Psi_{\bf k}^{\dagger}
\mathcal H_{\bf k}({\bf Q}_{\rm s})
\Psi_{\bf k}
+
\frac{|\Delta_{\rm s}|^2}{g_{\rm SC}}
+
\frac{|\Delta_{\rm e}|^2}{g_{\rm EI}}
+
\frac{\hbar^2 n_c}{2m_c}Q_{\rm s}^2 .
\end{equation}
Here \(n_c\) is the conduction-band electron density. The Hamiltonian  matrix is
\begin{equation}
\mathcal H_{\bf k}({\bf Q}_{\rm s})
=
\mathcal H_{\bf k}^{(0)}
+
Q_{{\rm s},i}v_{c,{\bf k},i}\Lambda,
\end{equation}
where
\begin{equation}
\mathcal H_{\bf k}^{(0)}
=
\begin{pmatrix}
\xi_{c,{\bf k}}
&
-|\Delta_{\rm s}|
&
-|\Delta_{\rm e}|
&
0
\\
-|\Delta_{\rm s}|
&
-\xi_{c,{\bf k}}
&
0
&
|\Delta_{\rm e}|
\\
-|\Delta_{\rm e}|
&
0
&
\xi_{v,{\bf k}+{\bf q}}
&
0
\\
0
&
|\Delta_{\rm e}|
&
0
&
-\xi_{v,-{\bf k}-{\bf q}}
\end{pmatrix},\quad\quad
\Lambda
=
\begin{pmatrix}
1&0&0&0\\
0&1&0&0\\
0&0&0&0\\
0&0&0&0
\end{pmatrix}.
\end{equation}
In this convention, the phase twist couples only to the conduction-band sector because only the conduction electrons participate directly in the SC pairing.

The corresponding bare Matsubara Green function is~\cite{mahan2013many,abrikosov2012methods}
\begin{equation}
G({\bf k},i\omega_m)
=
\left[
i\omega_m
-
\mathcal H_{\bf k}^{(0)}
\right]^{-1}.
\end{equation}
Then, the free energy in the presence of the phase twist is
\begin{equation}
\Omega({\bf Q}_{\rm s})
=
\Omega(0)
+
\frac{\hbar^2n_c}{2m_c}Q_{\rm s}^2
-
T
\sum_{{\bf k},m}
\ln
\det
\left[
1-
G({\bf k},i\omega_m)
Q_{{\rm s},i}v_{c,{\bf k},i}\Lambda
\right],
\end{equation}
The superfluid density tensor is defined as~\cite{schrieffer1964theory,abrikosov2012methods}
\begin{equation}
\rho_{s,ij}
=\frac{m_c}{\hbar^2}\times
\left.
\frac{\partial^2\Omega({\bf Q}_{\rm s})}
{\partial Q_{{\rm s},i}\partial Q_{{\rm s},j}}
\right|_{{\bf Q}_{\rm s}=0}.
\label{rho_definition}
\end{equation}
Expanding the free energy to quadratic order in \({\bf Q}_{\rm s}\), we obtain
\begin{equation}
\rho_{s,ij}
=
K^{\rm dia}_{ij}
+
K^{\rm para}_{ij},
\end{equation}
where the diamagnetic contribution is
\begin{equation}
K^{\rm dia}_{ij}
=
n_c\delta_{ij},
\end{equation}
and the paramagnetic contribution is given by the static current-current correlation function~\cite{abrikosov2012methods,schrieffer1964theory,mahan2013many},
\begin{equation}
K^{\rm para}_{ij}
=
-\frac{m_c}{\hbar^2}
\frac{1}{\beta}
\sum_{{\bf k},i\omega_m}v_{c,{\bf k},i}v_{c,{\bf k},j}
{\rm Tr}
\left[
G({\bf k},i\omega_m)
\Lambda
G({\bf k},i\omega_m)
\Lambda
\right].
\label{Kpara}
\end{equation}

Following the derivations in Sec.~\ref{sdsd}, using the unitary matrix \(U_{\bf k}\), the Green function can be written as
\begin{equation}
G({\bf k},i\omega_m)
=
U_{\bf k}
\frac{1}{
i\omega_m
-
{\rm diag}
(\epsilon_1,\epsilon_2,\epsilon_3,\epsilon_4)
}
U_{\bf k}^{\dagger}.
\end{equation}
Here, $
\epsilon_n({\bf k})
\in
\left\{
E_+({\bf k}),
-E_+({\bf k}),
E_-({\bf k}),
-E_-({\bf k})
\right\}$.  The current vertex in the quasiparticle basis is
\begin{equation}
\widetilde{\Lambda}({\bf k})
=
U_{\bf k}^{\dagger}
\Lambda
U_{\bf k}=\begin{pmatrix}
L_{11} & L_{12} & L_{13} & L_{14}
\\
L_{21} & L_{22} & L_{23} & L_{24}
\\
L_{31} & L_{32} & L_{33} & L_{34}
\\
L_{41} & L_{42} & L_{43} & L_{44}
\end{pmatrix},
\end{equation}
with
\begin{eqnarray}
L_{\lambda\lambda'}
&=&
\frac{
\Delta_s^2
(\xi_v^2-\epsilon_\lambda^2)
(\xi_v^2-\epsilon_{\lambda'}^2)
+
X_\lambda X_{\lambda'}
(\xi_v+\epsilon_\lambda)
(\xi_v+\epsilon_{\lambda'})
}{
\mathcal N_\lambda
\mathcal N_{\lambda'}
},
\end{eqnarray}
where 

\begin{equation}
X_\lambda
=
(\xi_c-\epsilon_\lambda)
(\xi_v-\epsilon_\lambda)
-
\Delta_e^2, \quad\quad
\mathcal N_\lambda=
\Big[
\Delta_s^2
(\xi_v^2-\epsilon_\lambda^2)^2
+
X_\lambda^2
(\xi_v+\epsilon_\lambda)^2
+
\Delta_e^2\Delta_s^2
(\xi_v+\epsilon_\lambda)^2
+
\Delta_e^2X_\lambda^2
\Big]^{1/2}.
\end{equation}
Substituting this spectral representation into the current-current correlation function gives
\begin{eqnarray}
K^{\rm para}_{ij}
&=&\frac{m_c}{\hbar^2}
\sum_{{\bf k},n,m}v_{c,{\bf k},i}v_{c,{\bf k},j}
\widetilde{\Lambda}^{nm}({\bf k})
\widetilde{\Lambda}^{mn}({\bf k})
\frac{1}{\beta}
\sum_{i\omega_m}
\frac{1}{
i\omega_m-\epsilon_n({\bf k})
}
\frac{1}{
i\omega_m-\epsilon_m({\bf k})
}\nonumber\\
&=&
\frac{m_c}{\hbar^2}\sum_{{\bf k},n,m}v_{c,{\bf k},i}v_{c,{\bf k},j}
\frac{
f(\epsilon_n)-f(\epsilon_m)
}{
\epsilon_n-\epsilon_m
}
\,
\widetilde{\Lambda}^{nm}({\bf k})
\widetilde{\Lambda}^{mn}({\bf k}).
\label{Kpara_band_explicit}
\end{eqnarray}
For \(n=m\), the ratio is understood as $
\frac{
f(\epsilon_n)-f(\epsilon_m)
}{
\epsilon_n-\epsilon_m
}
\rightarrow
f'(\epsilon_n)
=
-\beta f(\epsilon_n)
\left[
1-f(\epsilon_n)
\right]$.
Thus the final superfluid density tensor is
\begin{equation}
\rho_{s,ij}
=n_c\delta_{ij}+
\frac{m_c}{\hbar^2}
\sum_{{\bf k},n,m}v_{c,{\bf k},i}v_{c,{\bf k},j}
\frac{
f(\epsilon_n)-f(\epsilon_m)
}{
\epsilon_n-\epsilon_m
}
\,
\widetilde{\Lambda}^{nm}({\bf k})
\widetilde{\Lambda}^{mn}({\bf k}).
\label{rho_band_final}
\end{equation}

In the absence of EI order,
\begin{equation}
|\Delta_{\rm e}|
\rightarrow
0,
\end{equation}
the result reduces to the conventional BCS expression for a single parabolic band. In the coexistence phase, however, the finite-momentum EI order reconstructs the quasiparticle spectrum through the dispersions
\begin{equation}
\xi_{v,{\bf k}+{\bf q}},
\qquad
\xi_{v,-{\bf k}-{\bf q}}.
\end{equation}
Because the ordering vector \({\bf q}\) selects a preferred direction in momentum space, the quasiparticle spectrum ${\pm}E_{\pm}({\bf k})$ and current vertices $\widetilde{\Lambda}({\bf k})$ (in the quasiparticle basis) become anisotropic. Consequently, the superfluid density tensor generally satisfies
\begin{equation}
\rho_{s,xx}
\neq
\rho_{s,yy}.
\end{equation}
This anisotropic superfluid response provides a direct signature of finite-momentum EI order coexisting with superconductivity and naturally generates an emergent nematic SC state even in an otherwise isotropic electronic system.

\section{EI-pairing phase space in momentum space: the origin of EI-SC coexistence}

In this section, we discuss the underlying physical mechanism responsible for the coexistence between superconductivity and the finite-momentum EI density-wave order.

\subsection{FFLO-Like finite-${\bf q}$ EI pairings}

We start with the FFLO-like finite-${\bf q}$ electron-hole pairing and the corresponding EI-pairing phase space in momentum space. We first set $\Delta_s=0$ and focus only on the finite-momentum EI order that couples a conduction-electron state to a valence-hole state with a finite center-of-mass momentum ${\bf q}$, which optimizes the overlap between the electron and hole Fermi surfaces. The reduced mean-field Hamiltonian in the absence of superconductivity is
\begin{equation}
H_{\rm EI}
=
\sum_{{\bf k},\sigma}
\begin{pmatrix}
c_{{\bf k}\sigma}^{\dagger} &
v_{{\bf k}+{\bf q},\sigma}^{\dagger}
\end{pmatrix}
\begin{pmatrix}
\xi_{c,{\bf k}} & -\Delta_e({\bf q})\\
-\Delta_e^\ast({\bf q}) & \xi_{v,{\bf k}+{\bf q}}
\end{pmatrix}
\begin{pmatrix}
c_{{\bf k}\sigma}\\
v_{{\bf k}+{\bf q},\sigma}
\end{pmatrix}
+
\frac{|\Delta_e({\bf q})|^2}{g_{\rm EI}} .
\end{equation}
The diagonalization is performed by the transformation
\begin{equation}
\begin{pmatrix}
\alpha_{{\bf k}\sigma}\\
\beta_{{\bf k}\sigma}
\end{pmatrix}
=
\begin{pmatrix}
u_{\bf k} & -v_{\bf k}\\
v_{\bf k}^\ast & u_{\bf k}
\end{pmatrix}
\begin{pmatrix}
c_{{\bf k}\sigma}\\
v_{{\bf k}+{\bf q},\sigma}
\end{pmatrix},
\end{equation}
with coherence factors
\begin{equation}
u_{\bf k}^2
=
\frac{1}{2}
\Big(
1+
\frac{\delta_{\bf k}}
{\sqrt{\delta_{\bf k}^2+|\Delta_e|^2}}
\Big),
\qquad
v_{\bf k}^2
=
\frac{1}{2}
\Big(
1-
\frac{\delta_{\bf k}}
{\sqrt{\delta_{\bf k}^2+|\Delta_e|^2}}
\Big).
\end{equation}
The quasiparticle energies are
\begin{equation}
E_{\bf k}^{\pm}
=
\bar{\xi}_{\bf k}
\pm
\sqrt{
\delta_{\bf k}^2
+
|\Delta_e|^2
},
\end{equation}
where
\begin{equation}
\bar{\xi}_{\bf k}
=
\frac{
\xi_{c,{\bf k}}
+
\xi_{v,{\bf k}+{\bf q}}
}{2},
\qquad
\delta_{\bf k}
=
\frac{
\xi_{c,{\bf k}}
-
\xi_{v,{\bf k}+{\bf q}}
}{2}.
\end{equation}
The anomalous particle-hole expectation value is
\begin{equation}
\langle
v_{{\bf k}+{\bf q},\sigma}^{\dagger}
c_{{\bf k}\sigma}
\rangle
=
u_{\bf k}v_{\bf k}[
f(E_{\bf k}^{-})
-
f(E_{\bf k}^{+})],
\end{equation}
where $f(E)$ is the Fermi distribution function and the operator
$c_{{\bf k}\sigma}^{\dagger}
v_{{\bf k}+{\bf q},\sigma}$
creates an electron-hole pair carrying momentum ${\bf q}$.\\

The EI coherence is strongest in the momentum region where the conduction electron and valence hole around the Fermi surface are nearly degenerate,  i.e.,  near-perfect electron-hole band matching ($\bar\xi_{\bf k}\equiv0$), yielding 
 a fully paired EI state with as assumed in early EI theories. In this case, the lower branch is occupied and the upper branch is empty,
\begin{equation}
E_{\bf k}^{-}<0,
\qquad
E_{\bf k}^{+}>0.
\end{equation}
The corresponding BCS-like EI ground-state wave function is~\cite{PhysRev.158.462,PhysRevLett.19.439,RevModPhys.40.755}
\begin{equation}
|\Psi_{\rm EI}\rangle
=
\prod_{{\bf k},\sigma}
\left(
u_{\bf k}
+
v_{\bf k}
c_{{\bf k}\sigma}^{\dagger}
v_{{\bf k}+{\bf q},\sigma}
\right)
|0\rangle ,
\end{equation}
where $|0\rangle$ denotes the reference state with the valence band filled and the conduction band empty. 

In analogy with FFLO superconductivity~\cite{yang2018fulde,larkin1965zh,fulde1964superconductivity,Dong_2013,PhysRevA.89.013607,PhysRevLett.89.227002,PhysRevB.76.014522,PhysRevB.75.064511,PhysRevLett.114.110401,rqp1-jtcb}, a finite mismatch $\bar\xi_{\bf k}$ between the two Fermi surfaces can generate unpaired momentum regions.  These correspond to momenta for which one of the hybridized quasiparticle branches crosses the Fermi level, $
E_{\bf k}^{-}>0$ or $
E_{\bf k}^{+}<0$, 
yielding $
f(E_{\bf k}^{-})
-
f(E_{\bf k}^{+})
=0$, 
such that the EI correlation $\langle
v_{{\bf k}+{\bf q},\sigma}^{\dagger}
c_{{\bf k}\sigma}
\rangle
=
u_{\bf k}v_{\bf k}[
f(E_{\bf k}^{-})
-
f(E_{\bf k}^{+})]$ vanishes completely in these regions. Therefore, EI pairing becomes energetically unfavorable there. The ground state then contains both paired and unpaired sectors~\cite{fulde1964superconductivity},
\begin{equation}
|\Psi_{\rm EI}\rangle
=
|\Psi_{\rm paired}\rangle
\otimes
|\Psi_{\rm unpaired}\rangle .
\end{equation}
The paired sector is
\begin{equation}
|\Psi_{\rm paired}\rangle
=
\prod_{{\bf k}\in{\cal P},\sigma}
\left(
u_{\bf k}
+
v_{\bf k}
c_{{\bf k}\sigma}^{\dagger}
v_{{\bf k}+{\bf q},\sigma}
\right)
|0\rangle,\quad\quad
{\cal P}:~
E_{\bf k}^{-}<0<E_{\bf k}^{+}.
\end{equation}
The unpaired sector is determined by
\begin{equation}
|\Psi_{\rm unpaired}\rangle
=
\prod_{{\bf k}\in{\cal U}_c,\sigma}
c_{{\bf k}\sigma}^{\dagger}
\prod_{{\bf k}\in{\cal U}_v,\sigma}
v_{{\bf k}+{\bf q},\sigma}
|0\rangle , \qquad{\cal U}_c:
E_{\bf k}^{-}>0,
\qquad
{\cal U}_v:
E_{\bf k}^{+}<0.
\end{equation}
These unpaired regions, consisting of unpaired conduction-electron pockets and unpaired valence-hole pockets, are the particle-hole analog of the unpairing regions in FFLO superconductivity~\cite{yang2018fulde,larkin1965zh,fulde1964superconductivity,Dong_2013,PhysRevA.89.013607,PhysRevLett.89.227002,PhysRevB.76.014522,PhysRevB.75.064511,PhysRevLett.114.110401,rqp1-jtcb,PhysRevB.98.094507}.  The EI gap equation then becomes
\begin{equation}
\frac{1}{g_{\rm EI}}
=
\sum_{{\bf k}\in{\cal P}}
\frac{1}
{2\sqrt{\delta_{\bf k}^2+|\Delta_e|^2}},
\qquad
T=0.
\end{equation}
This equation shows explicitly that only momentum states in the paired sector  contribute to EI pairing. When the mismatch becomes too large, the unpaired region expands, the paired phase space shrinks continuously, and the EI gap is gradually suppressed to zero, leading to a continuous phase transition out of the EI state.  The finite vector ${\bf q}$ therefore plays the same role as the FFLO pairing momentum: it shifts the two Fermi surfaces relative to one another in momentum space and selects the region where particle-hole pairing is most efficient. The optimal ordering vector ${\bf q}$ is selected by maximizing the EI pairing phase space and therefore maximizing the EI gap amplitude,
\begin{equation}
{\bf q}
=
\arg\max_{\bf q}
|\Delta_e|.
\end{equation}
which is approximately equivalent to maximizing the overlap between the shifted conduction and valence Fermi surfaces,
\begin{equation}
{\bf q}
=
{\rm argmax}_{\bf q}
\int_{\rm FS}
d{\bf k}\,
\delta(\xi_{c,{\bf k}})
\delta(\xi_{v,{\bf k}+{\bf q}}).
\end{equation}
Therefore, within the present picture, both our calculations and experimental observations indicate that the magnitude of the ordering vector is approximately determined by
$
|{\bf q}|
\sim
k_{F,c}+k_{F,v}$, i.e.,  by the momentum-space separation between the conduction-electron and valence-hole pockets~\cite{x6k9-wgk9,px77-3gg1}.\\

\subsection{Ground-state many-body wavefunctions}

Importantly, the unpaired conduction-electron sector remains available for additional Cooper pairing~\cite{PhysRev.108.1175,schrieffer1964theory,tinkham2004introduction,abrikosov2012methods,mahan2013many}:
\begin{equation}
\prod_{{\bf k}\in{\cal U}_c,\sigma}
c_{{\bf k}\sigma}^{\dagger}
=
\prod_{{\bf k}\in{\cal U}_c}
c_{{\bf k}\uparrow}^{\dagger}
c_{-{\bf k}\downarrow}^{\dagger}
\;\;\Rightarrow\;\;
|\Psi_{\rm SC}\rangle
=
\prod_{{\bf k}\in{\cal U}_c}
\left(
u^{\rm SC}_{\bf k}
+
v^{\rm SC}_{\bf k}
c_{{\bf k}\uparrow}^{\dagger}
c_{-{\bf k}\downarrow}^{\dagger}
\right)
|0\rangle .
\end{equation}
Therefore, once a SC attractive interaction is introduced,  these residual conduction electrons can further condense into Cooper pairs, providing an additional SC condensation energy that further lowers the free energy.

In this picture, the paired EI sector and the residual SC sector occupy different regions of momentum space, allowing the two orders to coexist microscopically. \\

Consequently, one can approximately construct the coexistence ground state as
\begin{equation}\label{ESGS}
|\Psi_{\rm EI\text{-}SC}\rangle
=
\prod_{{\bf k}\in{\cal U}_c}
\left(
u^{\rm SC}_{\bf k}
+
v^{\rm SC}_{\bf k}
c_{{\bf k}\uparrow}^{\dagger}
c_{-{\bf k}\downarrow}^{\dagger}
\right)
\prod_{{\bf k}\in{\cal U}_v,\sigma}
v_{{\bf k}+{\bf q},\sigma}
\prod_{{\bf k}\in{\cal P},\sigma}
\left(
u_{\bf k}
+
v_{\bf k}
c_{{\bf k}\sigma}^{\dagger}
v_{{\bf k}+{\bf q},\sigma}
\right)
|0\rangle . 
\end{equation}
In this situation, the competition between EI pairing and SC pairing in momentum space becomes smooth, and consequently, the two orders can redistribute their pairing phase spaces continuously in momentum space, thereby enabling a continuous evolution and robust coexistence of the EI and SC orders.

\subsection{Absence of FFLO-type  superconductivity}
\label{SSECAFFLO}

As seen from the constructed ground state Eq.~(\ref{ESGS}) in the coexistence regime, while the EI order develops a finite momentum ${\bf q}$ to connect the conduction-electron and valence-hole pockets, the SC pairing still occurs between conduction-band states,
\begin{equation}
({\bf k}\uparrow,-{\bf k}\downarrow),
\end{equation}
as the leading SC instability continues to occur at zero center-of-mass momentum (BCS-type pairing), rather than forming an FFLO-type SC state with finite pairing momentum. 

{While the finite-${\bf q}$ EI order reconstructs the Fermi surface, it does not induce any additional instability toward finite-momentum superconductivity. The leading SC instability remains the conventional zero-$q$ SC state within the electron valley/pocket. 
Theoretically, the finite-${\bf q}$ EI  order belongs to the particle-hole channel, where the electron and hole Fermi surfaces can be mismatched. In this case, a finite center-of-mass momentum of the particle-hole pair can compensate  for the electron-hole Fermi-surface mismatch and enlarges the phase space available for EI pairing to reduce the free energy. {By contrast, the SC order considered here is the conventional spin-singlet particle-particle pairing channel. Since the EI order is spin independent and preserves time-reversal symmetry, it does not lift the spin degeneracy within the  electron valley/pocket. Consequently, the normal-state dispersion satisfies
\begin{equation}
\xi_{c,{\bf k}\uparrow}
=
\xi_{c,-{\bf k}\downarrow},
\end{equation}
so that the two states forming a Cooper pair remain exactly degenerate, without any momentum-space mismatch. Therefore, the conventional zero-$q$ Cooper channel involving $({\bf k},\uparrow)$ and $(-{\bf k},\downarrow)$ within the electron valley/pocket provides the maximum pairing phase space.} As for a more general SC pairing ansatz allowing finite center-of-mass momentum, 
$({\bf k}+{\bf q},\uparrow)
\leftrightarrow
(-{\bf k}+{\bf q},\downarrow)$. 
For ${\bf q}\neq0$, the two pairing states generally satisfy
\begin{equation}
\xi_{c,{\bf k}+{\bf q},\uparrow}
\neq
\xi_{c,-{\bf k}+{\bf q},\downarrow},
\end{equation}
which reduces the phase space available for Cooper pairing and directly suppresses the pairing susceptibility, thereby raising the SC free energy. Thus, the SC free energy is minimized at ${\bf q}=0$. This is also consistent with the conventional FFLO mechanism~\cite{yang2018fulde,larkin1965zh,fulde1964superconductivity,Dong_2013,PhysRevA.89.013607,PhysRevLett.89.227002,PhysRevB.76.014522,PhysRevB.75.064511,PhysRevLett.114.110401,rqp1-jtcb}, where finite-momentum superconductivity is stabilized only in the presence of a finite Fermi-surface mismatch between the pairing states. Experimentally, superconductivity in the relevant candidate systems (e.g., MoTe$_2$, see Refs.~\cite{x6k9-wgk9}) is also found to be consistent with Cooper pairing between states $({\bf k},\uparrow)$ and $(-{\bf k},\downarrow)$ within individual electron valleys/pockets. Beyond the intrinsic momentum $2{\bf K}_c$ associated with the valley center ${\bf K}_c$, no additional finite-${\bf q}$ SC component has been reported in these systems, which is fully consistent with our theoretical analysis.}

The emergence of nematic superconductivity in the coexistence regime originates entirely from the momentum-space competition between the EI and SC pairing channels. Since the finite-${\bf q}$ EI pairing predominantly occupies the momentum regions connected by ${\bf q}$, the remaining SC pairing phase space becomes intrinsically anisotropic. As a result, although the SC pairing itself still occurs between two states with zero center-of-mass momentum, the SC condensate inherits the rotational-symmetry breaking generated by the finite-${\bf q}$ EI pairing structure, yielding a pronounced $C_2$-anisotropy in the SC pairing phase space and superfluid transport, and hence giving rise to nematic superconductivity.

\section{Material-specific extensions: Real-space structure of possible density waves and experimental signatures}

 A finite-momentum EI order generically gives rise to an accompanying charge-density modulation, with the corresponding ordering wave vector expected to take the form ${\bf Q}_{\rm CDW}=\Delta{\bf K}+{\bf q}$. Here we briefly discuss the resulting real-space structure of this charge-density modulation, as well as possible pair-density-wave (PDW)-like phenomena associated with intravalley SC pairing in the underlying valley structure.

For candidate systems such as NaAlSi~\cite{qfqm-qp82,PhysRevB.81.245114,Yi2019NaAlSi,Muechler2019NaAlSi} and monolayer 1T$'$-MoTe$_2$~\cite{x6k9-wgk9,px77-3gg1,Rhodes2021}, the conduction-band electron pockets responsible for superconductivity are approximately isotropic, while the anisotropy primarily originates from the EI sector through either anisotropic hole pockets or intervalley structure. Consequently, we call for experimental verification of the predicted nematic SC response in these candidate materials. In particular, within the coexistence regime [i.e., once the finite-$q$ EI (excitonic CDW) correlation develops], the SC transport is  expected to be enhanced predominantly along the direction perpendicular to the EI ordering vector ${\bf q}$, leading to a pronounced $C_2$-anisotropy despite the nearly isotropic conduction band. Upon further electrostatic gating or carrier doping into the superconductivity-dominated regime where the EI order is fully suppressed, the rotational symmetry should be restored, and the SC response is expected to recover an approximately isotropic character.

\subsection{Analysis of the EI charge-density-wave  order}
{
Here we analyze the structure of the EI CDW  order when the conduction and valence bands are centered at different valleys. In this situation, the EI order does not merely carry a small residual momentum, but also contains the large intervalley momentum connecting the two band extrema. Let the conduction-band minimum be located near valley momentum ${\bf K}_c$, and the valence-band maximum be located near ${\bf K}_v$. We define the intervalley vector
\begin{equation}
\Delta{\bf K}
=
{\bf K}_c
-
{\bf K}_v .
\end{equation}
The EI order parameter couples a conduction electron near ${\bf K}_c$ to a valence hole near ${\bf K}_v$. Consequently, the physical ordering wave vector contains both the intervalley momentum $\Delta{\bf K}$ and an additional small residual nesting vector ${\bf q}$ that optimizes the remaining Fermi-surface mismatch,
\begin{equation}
{\bf Q}_{\rm CDW}
=
\Delta{\bf K}
+
{\bf q}.
\end{equation}
The corresponding real-space interband coherence takes the form~\cite{Haug1994,kittel1963quantum,PhysRevB.97.224423,PhysRevLett.35.120,gruner1988dynamics,gruner1985charge,PhysRevLett.134.066602} 
\begin{equation}
\left\langle
v_{\sigma}^{\dagger}({\bf r})
c_{\sigma}({\bf r})
\right\rangle
=
\Phi_e({\bf q})
e^{i{\bf Q}_{\rm CDW}\cdot{\bf r}},
\end{equation}
where $\Phi_e({\bf q})$ is a complex amplitude determined by the residual finite-momentum pairing instability. The EI condensate therefore induces a real-space density modulation~\cite{gruner1988dynamics,gruner1985charge,PhysRevLett.134.066602} 
\begin{eqnarray}
\rho({\bf r})
&=&
\rho_0
+
\rho_{\bf q}
e^{i{\bf Q}_{\rm CDW}\cdot{\bf r}}
+
\rho_{-{\bf q}}
e^{-i{\bf Q}_{\rm CDW}\cdot{\bf r}}=
\rho_0
+
|\rho_{\bf q}|
\cos({\bf Q}_{\rm CDW}\cdot{\bf r}+\theta),
\end{eqnarray}
where
\begin{equation}
\rho_{\bf q}
=
|\rho_{\bf q}|
e^{i\theta}
\propto
\Delta_e({\bf q}).
\end{equation}
Thus, the EI condensate simultaneously realizes a CDW state with physical ordering vector
\begin{equation}
{\bf Q}_{\rm CDW}
=
\Delta{\bf K}
+
{\bf q}.
\end{equation}
Physically, the large intervalley vector $\Delta{\bf K}$ is primarily fixed by the underlying band geometry and lattice structure, while the smaller residual vector ${\bf q}$ adjusts the remaining Fermi-surface mismatch between the electron and hole pockets. Changes in the band overlap $E_g$, carrier density, strain, or effective masses modify the relative sizes and shapes of the electron and hole Fermi surfaces and therefore mainly tune the residual nesting vector ${\bf q}$. This distinction between ${\bf Q}_{\rm CDW}$ and ${\bf q}$ is important for interpreting experiments. Scattering probes such as X-ray diffraction or STM measurements observe the physical density-wave modulation vector ${\bf Q}_{\rm CDW}=\Delta{\bf K}+{\bf q}$, whereas the low-energy EI instability itself is governed primarily by the residual nesting vector ${\bf q}$.}

\subsubsection{Pinning mechanism of the nematic axis in  monolayer 1T'-MoTe$_2$}
\label{SMMMT}

In monolayer 1T$'$-MoTe$_2$, the large intervalley momentum separation $
\Delta{\bf K}
=
{\bf K}_c-{\bf K}_v
=(0,\frac{\pi}{3b})$ [where $b$
is the lattice constant along the zigzag Mo chains], fixed by the underlying lattice symmetry, 
provides the dominant momentum component of the translational-symmetry-breaking order, and the remaining finite-momentum mismatch between the electron and hole pockets after they are aligned to a common momentum center is described by an additional vector ${\bf q}$.  When the preferred ordering vector approaches a rational fraction of a reciprocal lattice vector, higher-order Umklapp processes generate additional locking terms in the free energy. Such commensurability effects can stabilize particular ordering-vector directions and reduce the continuous rotational degeneracy present in the isotropic continuum limit, so that the EI ordering vector ${\bf q}$ tends to align parallel to $\Delta{\bf K}$,
\begin{equation}
{\bf q}\parallel \Delta{\bf K},
\end{equation}
i.e., optimal ${\bf q}$ is naturally chosen to be parallel to $\Delta{\bf K}$.   Consequently, the CDW vector remains pinned along the $\Delta{\bf K}$ direction,
\begin{equation}
{\bf Q}_{\rm CDW}\parallel \Delta{\bf K}.
\end{equation}
This is consistent with experiments on monolayer 1T$'$-MoTe$_2$, which report a density-wave modulation vector approximately given by~\cite{x6k9-wgk9,px77-3gg1}
\begin{equation}
{\bf Q}^{\rm exp}_{\rm CDW}
\simeq
\left(
0,\frac{2\pi}{5b}
\right)\parallel\Delta{\bf K}=\left(0,\frac{\pi}{3b}\right),
\end{equation}
i.e., the experimentally observed vector ${\bf Q}^{\rm exp}_{\rm CDW}$ connects the electron and hole Fermi-surface pockets rather than the valley centers themselves. This indicates that the actual ordering vector is not determined solely by the intervalley separation between the valley centers. Instead, its dominant component is fixed by $\Delta{\bf K}$, while the smaller correction ${\bf q}$ compensates the residual Fermi-surface mismatch and points  parallel to $\Delta{\bf K}$. It should also be noted that, as discussed in Sec.~\ref{secLO}, the full LO-like EI state contains both ${\bf Q}=\Delta{\bf K}+{\bf q}$ and its time-reversal partner $-{\bf Q}$. Therefore, the energetic tendency discussed here applies equally to both sectors and does not modify the resulting conclusions here.

Within the  proposed mechanism, such a finite-momentum EI modulation selectively suppresses EI pairing predominantly along the corresponding momentum direction, while leaving the orthogonal momentum sectors available for Cooper pairing. Consequently, we expect the SC response and superfluid stiffness to become enhanced mainly along the direction perpendicular to the modulation vector ${\bf Q}^{\rm exp}_{\rm CDW}$ in the coexistence regime. This provides a direct experimentally testable signature of the proposed EI-SC coexistence mechanism and the associated nematic superconductivity.

\begin{figure}[b]
\centering
\begin{minipage}[c]{0.42\linewidth}
    \centering
    \includegraphics[width=\linewidth]{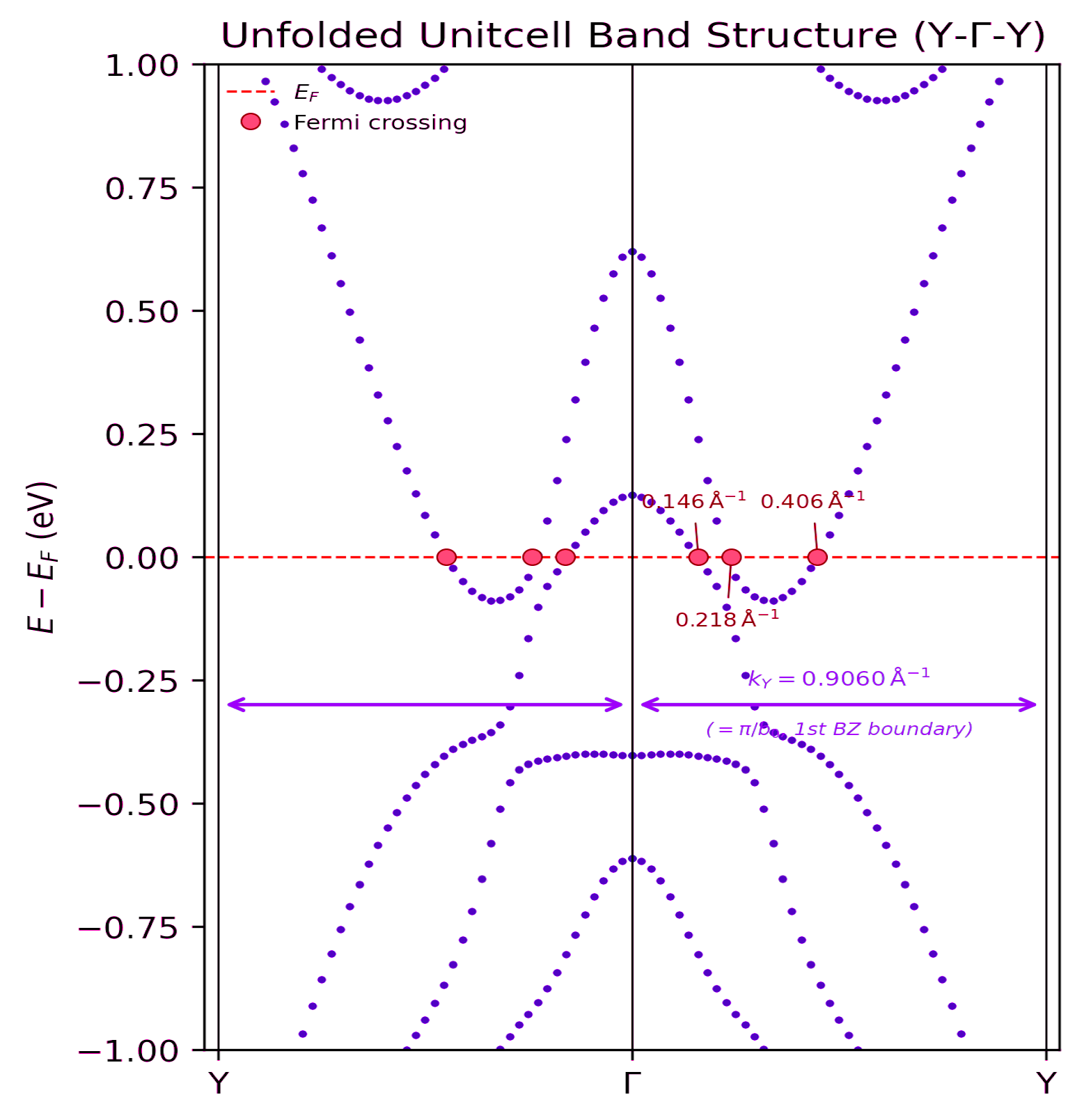}
\end{minipage}
\hfill
\begin{minipage}[c]{0.57\linewidth}
    \caption{
    DFT band structure of monolayer MoTe$_2$.
    First-principles calculations were performed using the
    projector augmented-wave (PAW) method~\cite{blochl1994projector}
    as implemented in the Vienna \textit{Ab initio} Simulation
    Package (VASP)~\cite{kresse1993abinitio}.
    The exchange-correlation functional was treated within
    the generalized gradient approximation (GGA) using the
    Perdew--Burke--Ernzerhof (PBE) parametrization~\cite{perdew1996generalized}.
    Monolayer MoTe$_2$ was modeled using a slab geometry,
    with spin-orbit coupling (SOC) included self-consistently.
    Maximally localized Wannier functions were constructed
    from the SOC-included electronic structure using
    \textsc{Wannier90}~\cite{mostofi2008wannier90},
    with initial projections onto Mo $d$ and Te $p$ orbitals.
    }
    \label{SuppFig_DFT}
\end{minipage}

\end{figure}

Following the discussion above, in monolayer MoTe$_2$ with $\Delta{\bf K}=(0,\frac{\pi}{3b})$, a nonzero finite-momentum particle-hole coherence necessarily
produces a charge-density modulation at the ordering wave vector
\begin{equation}
{\bf Q}_{\rm CDW}
=\Delta{\bf K}+{\bf q},
\end{equation}
and as numerically demonstrated in our work, to maximize the phase space available for electron-hole coherence, the excitonic instability generally favors a finite momentum shift that compensates the mismatch between the electron and hole Fermi surfaces, approximately given by
\begin{equation}
{\bf q}
\simeq
(k_{F,v}-k_{F,c})\hat{\bf y},
\end{equation}
where $k_{F,c}$ and $k_{F,v}$ denote the Fermi wavevectors of the conduction and valence pockets, respectively, measured from the corresponding pocket centers.
Then, according to a density functional
theory (DFT) calculation,  one finds 
$\Delta{\bf K}=(0,0.302~\mathrm{\AA}^{-1})$ and hence 
${\bf Q}_{\rm CDW}
=\Delta{\bf K}+{\bf q}=(0,0.218~\mathrm{\AA}^{-1}+0.146~\mathrm{\AA}^{-1})=(0,0.364~\mathrm{\AA}^{-1})
\approx\Big(
0,\frac{2\pi}{5b}
\Big)$. Then, the experimentally observed ordering vector
${\bf Q}^{\rm exp}_{\rm CDW}$ naturally emerges according to the mechanism proposed in this work.

\subsubsection{Pinning mechanism of the nematic axis in square-net semimetal NaAlSi}
\label{sssecnaalsi}

In the square-net semimetal NaAlSi~\cite{qfqm-qp82,PhysRevB.81.245114,Yi2019NaAlSi,Muechler2019NaAlSi}, the situation is qualitatively different because $\Delta{\bf K}=0$, and hence, 
\begin{equation}
{\bf Q}_{\rm CDW}={\bf q}.
\end{equation}
The EI ordering vector ${\bf q}$ is determined primarily by the anisotropic electron-hole Fermi-surface mismatch.   In the square-net semimetal NaAlSi~\cite{qfqm-qp82,PhysRevB.81.245114,Yi2019NaAlSi,Muechler2019NaAlSi}, the conduction-band electron pocket responsible for superconductivity is approximately isotropic, while the valence-band hole pocket is strongly anisotropic. In this situation, the dispersions may be approximated as
\begin{equation}
\xi_c({\bf k})
=
-\frac{E_g}{2}
+
\frac{k^2}{2m_c}
-
\mu,\qquad
\xi_v({\bf k})
=
\frac{E_g}{2}
-
\frac{k_x^2}{2m_{v,x}}
-
\frac{k_y^2}{2m_{v,y}}
-
\mu.
\end{equation}
The hole-pocket Fermi surface is therefore elliptic rather than circular~\cite{qfqm-qp82}. Consequently, the approximate nesting condition
\begin{equation}
\xi_c({\bf k})
\simeq
\xi_v({\bf k}+{\bf q})
\end{equation}
is satisfied anisotropically in momentum space. Since in NaAlSi the conduction-band Fermi momentum is either smaller than or more closely matched to the short-axis Fermi momentum of the elliptic hole pocket~\cite{qfqm-qp82}, the electron-hole mismatch is minimized along the short-axis direction of the hole pocket. Thus, the EI pairing phase space is maximized for ${\bf q}$ aligned along the short-axis direction, thereby naturally pinning the nematic axis through the anisotropic hole pocket, consistent with experimental observations that the nematic response is locked to the crystalline axis in NaAlSi~\cite{qfqm-qp82}. 

The SC response is then enhanced predominantly along the orthogonal direction, yielding a pinned nematic SC axis in NaAlSi.

\subsection{Possible valley-structure-induced PDW-like gap modulations}

As discussed in Sec.~\ref{SSECAFFLO}, the SC pairing considered in the present work does not correspond to an FFLO SC state, since the Cooper pairing carries zero center-of-mass momentum within the low-energy continuum theory. Even if the conduction band is centered around finite valley momenta $\pm{\bf K}_c$ and the SC order parameter in real space formally takes the form
\begin{eqnarray}
|\Delta_{\rm SC}({\bf r})|
&\sim&
e^{i2{\bf K}_c\cdot{\bf r}}
\left\langle
c_{\downarrow}({\bf r})
c_{\uparrow}({\bf r})
\right\rangle
+
e^{-i2{\bf K}_c\cdot{\bf r}}
\left\langle
c_{\downarrow}({\bf r})
c_{\uparrow}({\bf r})
\right\rangle.
\end{eqnarray}
typically, however, crystal momentum is defined modulo a reciprocal lattice vector ${\bf G}$. Consequently, the SC condensate remains translationally invariant provided
\begin{equation}
2{\bf K}_c
=
{\bf G}.
\end{equation}
This condition is satisfied in many multivalley systems, such as monolayer transition-metal dichalcogenides (e.g., MoS$_2$, WS$_2$, WSe$_2$) and graphene-based systems, where the valleys are located at symmetry-related momenta. In such situations, the Cooper pair carries zero crystal momentum modulo the lattice periodicity, and the SC state therefore remains an ordinary uniform superconductor.

However, in systems such as monolayer 1T$'$-MoTe$_2$~\cite{x6k9-wgk9,px77-3gg1,Rhodes2021}, where $
{\bf K}_c
=(
0,
\frac{\pi}{3b})$, 
one obtains
\begin{equation}
2{\bf K}_c
=
\left(
0,
\frac{2\pi}{3b}
\right)
\not\equiv0
\quad
\left(
\mathrm{mod}\ 
{\bf G}_b
=
\left(
0,
\frac{2\pi}{b}
\right)
\right),
\end{equation}
which becomes experimentally equivalent to a modulation wave vector $
{\bf Q}_{P}^{b}
\simeq(
0,
\frac{2\pi}{3b})$, in qualitative agreement with recent STM experiments on monolayer 1T$'$-MoTe$_2$ reporting SC gap modulations with an approximately $3b$ real-space period along the zigzag Mo-chain direction~\cite{x6k9-wgk9,px77-3gg1,Rhodes2021}. {Nevertheless, it should be emphasized that while the unique valley structure in monolayer 1T$'$-MoTe$_2$ generates lattice-scale PDW-like modulations of the SC  gap, as realized in previous studies~\cite{x6k9-wgk9,px77-3gg1,Rhodes2021}, this effect originates purely from the lattice-periodic embedding of the valley momenta into crystal momentum space and does not involve a corresponding shift of the low-energy quasiparticle group velocities or superfluid center-of-mass motion. Consequently, the resulting modulation does not by itself produce nematic SC responses or anisotropic superfluid transport, and is not expected to qualitatively modify the main conclusions of the present work. The SC transport remains approximately isotropic in the SC-dominated regime, where the competing EI order is absent, despite the appearance of real-space gap modulations at the lattice scale. By contrast, within the SC-EI coexistence regime, the finite-momentum EI order yields a genuine nematic SC response characterized by strongly anisotropic superfluid transport.}\\

\begin{figure}[h]
  {\includegraphics[width=8.6cm]{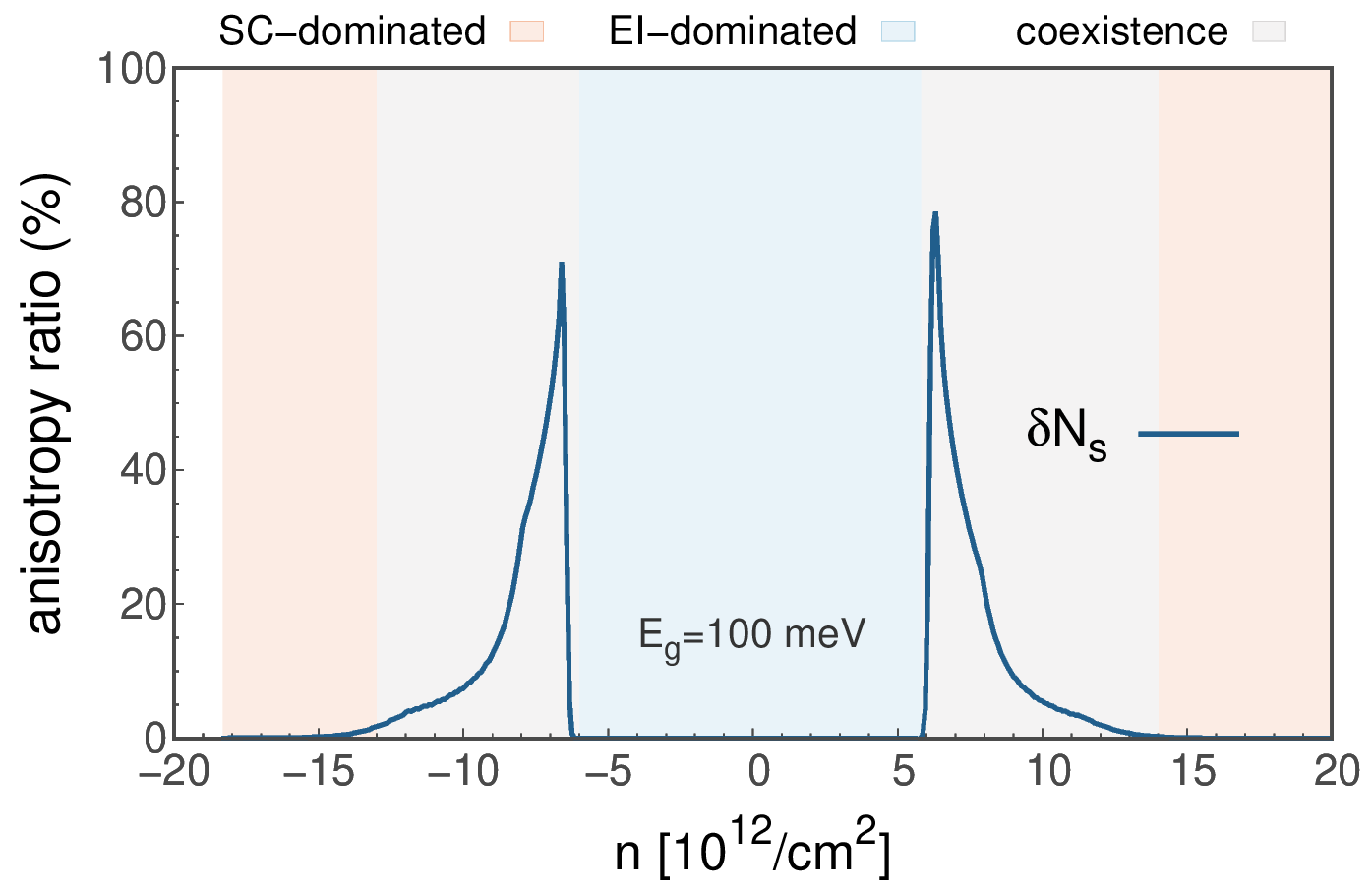}}
\caption{SC nematicity characterized by the anisotropy ratio $\delta N_s$ as functions of the net carrier density
$n=n_c-n_v$. The phase regimes are determined from Fig.~2(a) in the main text.
}
\label{SFIG2}
\end{figure}

\section{Additional results}

\subsection{SC nematicity and significant anisotropy ratio}

To further characterize the anisotropic SC state, we evaluate the superfluid-density anisotropy ratio
\begin{equation}
\delta N_s=
\frac{\rho_{s,xx}-\rho_{s,yy}}
{\rho_{s,xx}+\rho_{s,yy}},
\end{equation}
where $\rho_{s,xx}$ and $\rho_{s,yy}$ denote the superfluid stiffness perpendicular and parallel to the ordering wave-vector axis ${\hat {\bf q}}$, respectively. As shown in Fig.~\ref{SFIG2}, a pronounced SC nematicity develops over a broad doping range within the SC-EI coexistence regime.

The anisotropy ratio reaches values as large as $\sim70\%$ on both the hole- and electron-doped sides within the SC-EI coexistence regime, demonstrating a highly anisotropic SC condensate. Such a strong anisotropy indicates that the SC phase stiffness becomes strongly suppressed along the ordering direction, while remaining comparatively robust in the transverse direction. This behavior suggests the emergence of a quasi-one-dimensional SC tendency induced by the finite-momentum EI CDW background. In contrast, once the system enters the SC-dominated regime and the EI order is suppressed, the superfluid response rapidly becomes nearly isotropic. The sharp crossover from strongly anisotropic to nearly isotropic superfluid transport therefore provides a clear and experimentally accessible signature of the SC-EI coexistence phase.

\subsection{Temperature dependence}

\begin{figure}[h]
    \centering
    \includegraphics[width=0.95\linewidth]{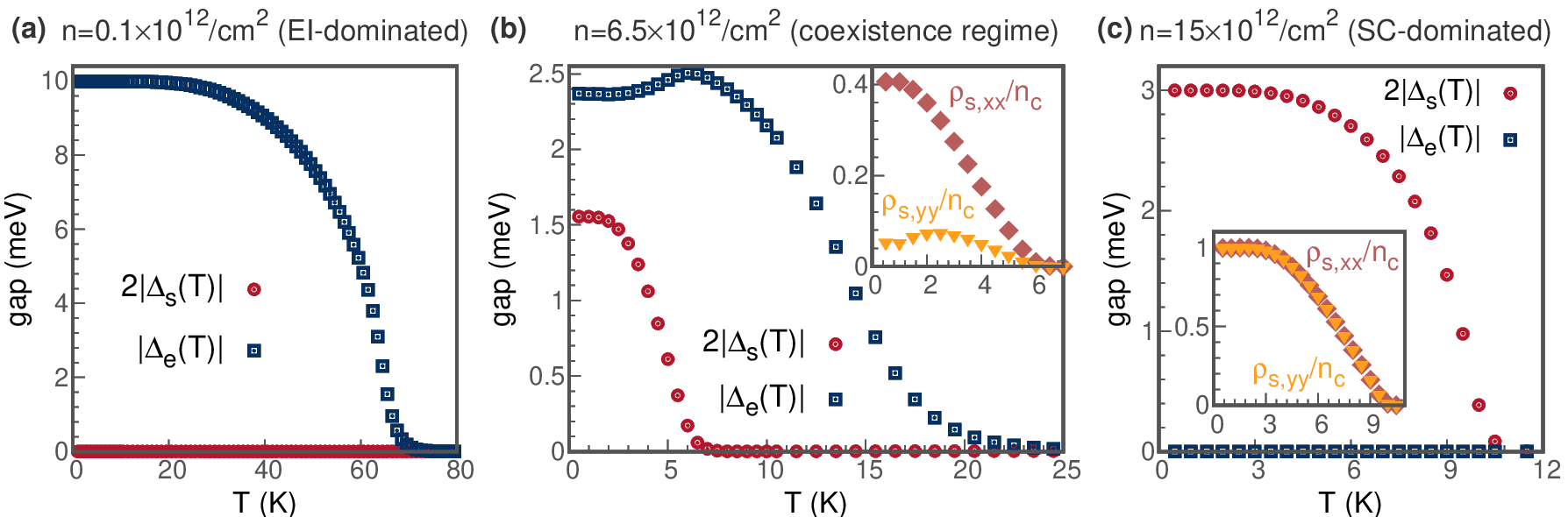}
\caption{{
Temperature dependence of the EI gap $|\Delta_e|$ and SC gap $|\Delta_s|$ at representative carrier densities obtained from the mean-field calculations. The insets in (b) and (c) shows the temperature dependence of the superfluid-density tensor components.  $E_g=100~$meV.}}
\label{figR4}
\end{figure}

{In this part, we present the temperature evolution within the mean-field framework, including the temperature dependence of $|\Delta_e|$, $|\Delta_s|$, and the superfluid-density anisotropy at three representative carrier densities. As shown in Fig.~\ref{figR4}(a) [Fig.~\ref{figR4}(c)], in the EI-dominated (SC-dominated) regime, the EI gap $|\Delta_e(T)|$ (SC gap $|\Delta_s(T)|$) exhibits the characteristic mean-field temperature dependence, as expected. In the coexistence regime, the temperature evolution of the gaps shows distinct behaviors due to the competition between the EI and SC orders. In particular, as temperature increases and the SC gap is rapidly suppressed to zero, the EI gap exhibits a transient anomalous upturn, reflecting the competition between the EI and SC pairing channels. Moreover, a pronounced anisotropy of the superfluid-density tensor [inset of Fig.~\ref{figR4}(b)] appears within the temperature range where the SC gap remains finite in the coexistence regime, and disappears when the SC gap vanishes. In the SC-dominated regime, the superfluid-density tensor recovers rotational symmetry, and its temperature dependence follows the conventional BCS behavior.}

{It should be emphasized that the mean-field approach provides the foundation for understanding the low-temperature ordered states; therefore, we present our main results at $T=0.5$ K, corresponding to the regime where the EI and SC orders are well established. Determining the finite-temperature phase boundaries in a two-dimensional system is a more involved problem. In low-dimensional systems, fluctuation effects, including collective phase fluctuations associated with the Nambu-Goldstone mode and Berezinskii-Kosterlitz-Thouless (BKT) physics, can substantially modify the transition temperatures. A quantitative description of these effects requires a fluctuation-corrected theory, which represents a distinct problem beyond the scope of  the mean-field framework.}

\section{Simulation details}

In the present study, following the conventional BCS framework~\cite{PhysRev.108.1175,schrieffer1964theory,tinkham2004introduction,abrikosov2012methods,mahan2013many}, a Debye energy cutoff was introduced for the SC  pairing channel around the conduction-band Fermi surface,
\begin{equation}
|\xi_{c,\mathbf k}|<\omega_D,
\end{equation}
while the EI pairing channel was evaluated within an energy window of $0.6E_g$ around the Fermi surface. In both the SC and EI channels, when the cutoff exceeds the corresponding band edge, it is replaced by the associated Fermi energy.

The self-consistent mean-field equations for the SC  gap $|\Delta_s|$, EI gap $|\Delta_e|$, and EI ordering vector ${\bf q}$ were solved iteratively on a two-dimensional momentum grid parameterized by the conduction-band energy $\xi_{c,\mathbf k}$ and the momentum angle $\theta_{\bf k}$. Owing to the parabolic band structure, the valence-band dispersion $\xi_{v,\mathbf{k+q}}$ can then be mapped directly as a function of $\xi_{c,\mathbf k}$, $\theta_{\bf k}$, and ${\bf q}$,
\begin{eqnarray}
\xi_{v,\mathbf{k+q}}
&=&
-\frac{m_c}{m_v}\xi_{c,\mathbf k}
+
\frac{\hbar^2}{2m_v}
\left(
k_{F,v}^2
-
k_{F,c}^2
-
q^2
-
2kq\sin\theta_{\bf k}
\right),
\end{eqnarray}
where $\theta_{\bf k}$ is defined with respect to the ordering-vector direction ${\bf q}\parallel{\bf e}_y$. The carrier density was controlled by tuning the chemical potential $\mu$, which was determined self-consistently for each target density. In the numerical calculations, the EI ordering vector ${\bf q}$ was taken along the $y$ direction without loss of generality due to the rotational symmetry of the underlying isotropic model. The momentum summation was evaluated using a sufficiently dense momentum mesh to ensure numerical convergence of the SC and EI gap equations, thermodynamic potential, and superfluid density tensor. In the specific numerical calculations, the LO-like EI construction discussed in the main text was implemented explicitly, such that the momentum summation over the sector with $k_y>0$ was paired by ${\bf q}$, while the reversal-related sector with $k_y<0$ was simultaneously paired by $-{\bf q}$, thereby constructing the full LO-like EI state and avoiding double counting of pairing degrees of freedom.
 Equivalently, the explicit mean-field Hamiltonian and the associated gap equations are formulated within a single favorable pairing sector, while the contribution from the opposite sector is restored through the symmetry-related momentum summation.

The SC and EI gap equations were solved iteratively. Two different sets of initial trial configurations were employed in order to capture both the conventional uniform EI state and the FFLO-like finite-momentum EI state:
\begin{equation}
\Delta_s^{(0)},
\qquad
\Delta_e^{(0)},
\qquad
{\bf q}^{(0)}=0,\label{SS1}
\end{equation}
and
\begin{equation}
\Delta_s^{(0)},
\qquad
\Delta_e^{(0)},
\qquad
{\bf q}^{(0)}
=\pm
(k_{F,c}-k_{F,v}){\bf e}_y. \label{SS2}
\end{equation}
The order parameters were updated self-consistently according to
\begin{equation}
\Delta_s^{(n+1)}
=
g_{\rm SC}
\sum_{\bf k}
\rho_{\rm SC}
\big(
{\bf k},
\Delta_s^{(n)},
\Delta_e^{(n)}
\big),
\end{equation}
and
\begin{equation}
\Delta_e^{(n+1)}
=
g_{\rm EI}
\sum_{\bf k}
\rho_{\rm EI}
\big(
{\bf k},
\Delta_s^{(n)},
\Delta_e^{(n)}
\big).
\end{equation}
To stabilize the convergence, linear mixing was employed:
\begin{equation}
\Delta^{(n+1)}
\rightarrow
(1-\eta)\Delta^{(n)}
+
\eta \Delta_{\rm new}^{(n+1)},
\end{equation}
with a typical mixing parameter
\begin{equation}
\eta=0.22.
\end{equation}
For each iteration step, the ordering vector 
${\bf q}^{(n+1)}$ was determined by minimizing the thermodynamic potential using a steepest-descent optimization procedure. Starting from ${\bf q}^{(n)}$, the ordering vector was continuously relaxed until the thermodynamic potential reached a local minimum.

A small numerical broadening
\begin{equation}
\delta\sim10^{-4}~{\rm meV}
\end{equation}
was introduced in the quasiparticle denominators to improve numerical stability near gapless points. The reported results correspond to the lowest-free-energy solution obtained after full self-consistent convergence. Typical convergence criteria were
\begin{equation}
|\Delta_s^{(n+1)}-\Delta_s^{(n)}|
<
10^{-5}~{\rm meV},
\end{equation}
and
\begin{equation}
|\Delta_e^{(n+1)}-\Delta_e^{(n)}|
<
10^{-5}~{\rm meV}.
\end{equation}
Depending on the carrier density and microscopic parameters, the self-consistent iteration starting from the two initial configurations  [Eq.~(\ref{SS1}) and Eq.~(\ref{SS2})]  may converge either to only one nontrivial ordered state, with the other iteration becoming unstable or converging to a trivial solution, or to two distinct nontrivial self-consistent solutions. However, if both initial configurations converged to nontrivial self-consistent solutions, the corresponding thermodynamic potentials were compared, and the nontrivial solution with the lower free energy was selected as the final equilibrium state.\\

\begin{figure}[h]
    \centering
    \includegraphics[width=1.0\linewidth]{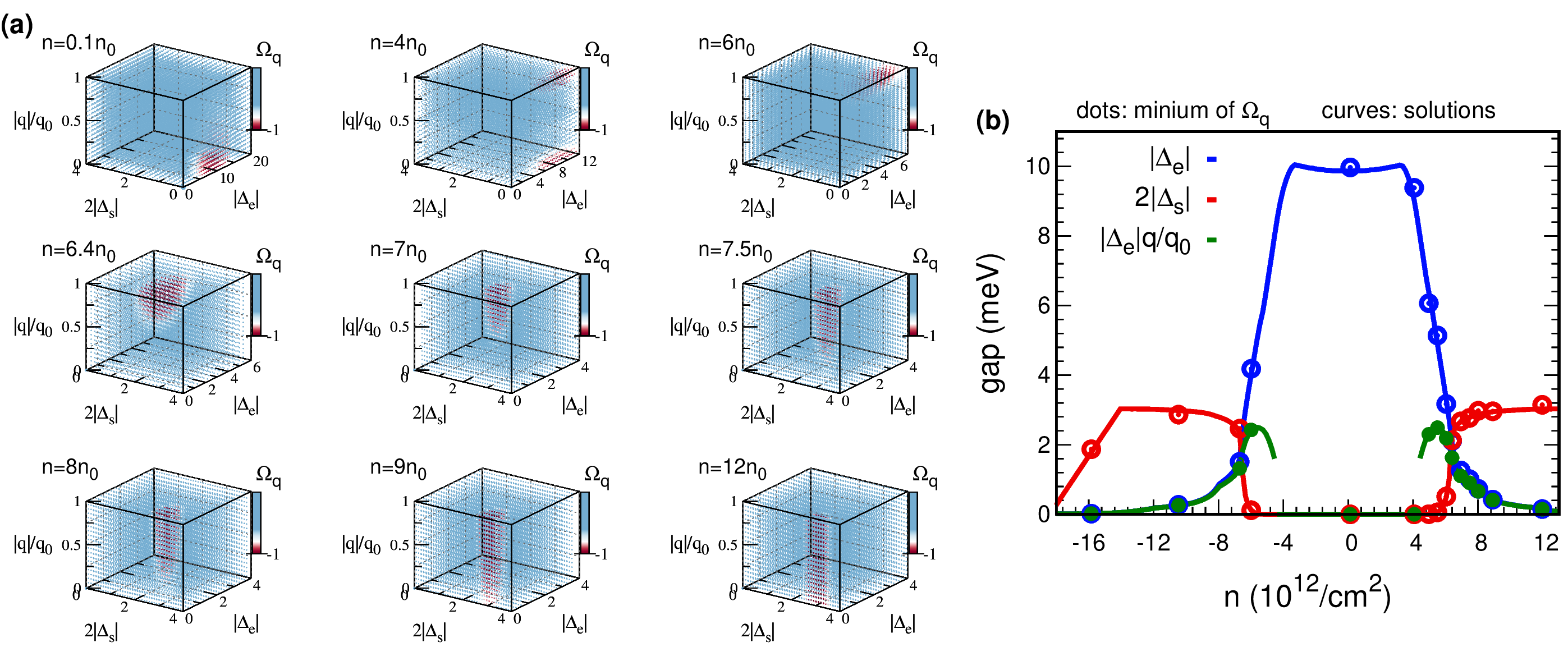}
\caption{
(a) Numerically calculated thermodynamic potential
$\Omega_{\bf q}(|\Delta_s|,|\Delta_e|,q{\bf e}_y)$  (shown in normalized units) 
at nine representative carrier densities covering
the EI-dominated, SC-dominated, and EI-SC coexistence regimes.
A double-well structure with two competing local minima
emerges around $n\sim4n_0$, reflecting the competition
between the nearly uniform $q=0$ and finite-$q$ EI states
at moderate electron-hole Fermi-surface mismatch.
As the carrier density further increases, the global minimum
switches from the $q=0$ branch to the finite-$q$ branch,
giving rise to a first-order transition (see Sec.~SI D), 
while at higher densities only a single minimum
at finite $q$ remains in the EI-SC coexistence regime. (b) Comparison between the global-minimum solutions
obtained from free-energy minimization and the
self-consistent solutions over a broader range of
carrier densities, showing good agreement, with small deviations arising from the numerical procedures employed in the two approaches. 
The mean absolute percentage deviations are
3.77\%, 3.61\%, and 2.48\% for
$|\Delta_s|$, $|\Delta_e|$, and $q$, respectively. 
$E_g=100~\mathrm{meV}$ and $T=0.5~\mathrm{K}$.
}
    \label{figSFS}
\end{figure}

{We have also performed
additional stability analyses by systematically scanning
the three-dimensional parameter space spanned by
$|\Delta_s|$, $|\Delta_e|$, and $q$,
thereby explicitly accounting for coupled variations
of the two gap amplitudes and the EI ordering wave vector. As shown in Fig.~\ref{figSFS}, we have considered
representative carrier densities covering the
EI-dominated, SC-dominated, and EI-SC coexistence
regimes. For all densities examined, we find that
the global minimum of the thermodynamic potential
$\Omega_{\bf q}(|\Delta_s|,|\Delta_e|,{\bf q})$
agrees well with the corresponding self-consistent
solution. These global-minimum solutions naturally
satisfy the local stability condition against coupled
variations of $|\Delta_s|$, $|\Delta_e|$, and $q$. These results support the energetic stability of the self-consistent solutions within the examined parameter space and the present theoretical framework. }

After obtaining the converged mean-field solution, the superfluid density tensor was evaluated numerically using
\begin{equation}
\rho_{s,ij}
=
n_c\delta_{ij}
+
\frac{m_c}{\hbar^2}
\sum_{{\bf k},n,m}
v_{c,{\bf k},i}
v_{c,{\bf k},j}
\frac{
f(\epsilon_n)-f(\epsilon_m)
}{
\epsilon_n-\epsilon_m
}
\,
\widetilde{\Lambda}^{nm}({\bf k})
\widetilde{\Lambda}^{mn}({\bf k}).
\label{rho_band_final_sim}
\end{equation}
In the practical numerical calculations, the gauge-invariance condition $
K^{\rm dia}_{ij}
=
-
K^{\rm para}_{ij}
(|\Delta_s|=0)$ 
was imposed to ensure the exact cancellation between the diamagnetic and paramagnetic responses in the normal state~\cite{schrieffer1964theory,abrikosov2012methods,PhysRevB.98.094507,PhysRevB.100.104513}.\\

This solving procedure is numerically demanding, since for each carrier density the coupled gap equations must be solved self-consistently while simultaneously optimizing the EI ordering vector ${\bf q}$ and evaluating the momentum-resolved quasiparticle spectrum. A more systematic exploration of the full phase diagram and its dependence on microscopic band parameters and temperature is beyond our  computational resources and scope of the present work and is left for future investigation.

\begin{table}[h]
\caption{
Parameters used in the self-consistent numerical calculations. Here $m_0$ denotes the bare electron mass and $D_c$ is the conduction-band density of states. The SC pairing interaction is restricted to the Debye window $|\xi_{c,\mathbf k}|<\omega_D$. For the parameter set used in the main text, the maximum EI gap obtained self-consistently is approximately $|\Delta_e|_{\rm max}\sim10~{\rm meV}$, while the maximum SC gap is approximately $|\Delta_s|_{\rm max}\sim1.8~{\rm meV}$. The calculations were performed in the regime $|\Delta_s|_{\rm max}\ll\omega_D$, ensuring that the SC  results are insensitive to the detailed high-energy structure of the cutoff.}
\label{tab:parameters}
\begin{tabular}{cc}
\hline
\hline
Parameter & Value \\
\hline
\hline
Temperature $T$ & $0.5~{\rm K}$ \\
Conduction-band effective mass $m_c$ & $0.42m_0$ \\
Valence-band effective mass $m_v$ & $0.44m_0$ \\
EI coupling $g_{\rm EI}D_c$ & $0.114$ \\
SC  coupling $g_{\rm SC}D_c$ & $0.062$ \\
Debye cutoff $\omega_D$ & $12~{\rm meV}$ \\
Band overlap $E_g$ & $20$--$100~{\rm meV}$ \\
\hline
\hline
\end{tabular}
\end{table}

{The effective masses of electrons and holes ($m_c=0.42m_0$ and $m_v=0.44m_0$) used in our model, as well as the SC pairing gap ($|\Delta_s(T=0)|=1.5$ meV in the SC-dominated regime), are chosen to be close to the values reported for monolayer 1T$'$-MoTe$_2$~\cite{x6k9-wgk9,px77-3gg1}. For the band overlap, we consider several representative values in the manuscript to more clearly illustrate the dependence of our results on this parameter. However, the most relevant microscopic energy scale associated with EI pairing has not yet been fully determined experimentally for monolayer 1T$'$-MoTe$_2$, owing to the limited microscopic characterization of this system. An important experimentally relevant hierarchy is that the EI energy scale is known to be much larger than the SC energy scale, as considered in our model.  However, the proposed mechanism is expected to be general for systems hosting competing excitonic and superconducting instabilities, characterized by a finite negative band overlap between the conduction and valence bands. It does not rely on a specific set of material parameters, but instead requires the coexistence between EI and SC orders.  Such a situation may naturally arise in candidate platforms including monolayer 1T$'$-MoTe$_2$ and NaAlSi, where excitonic (or excitonic CDW) behavior and superconductivity have both been reported experimentally. Moreover, in both systems, superconductivity is associated with conduction-band states with approximately isotropic dispersion.  We therefore expect that both material platforms may exhibit such an electrically
tunable evolution between nematic and isotropic superconductivity, enabling the control of SC transport anisotropy through external doping or gating, which calls for further experimental investigation.}

\bibliography{ref.bib}